\documentclass[runningheads]{llncs}

\usepackage[year=2024,ID=4870]{eccv}
\usepackage{eccvabbrv}

\usepackage{graphicx}
\usepackage{booktabs}
\usepackage{subcaption}
\usepackage{makecell}
\usepackage{multirow}
\usepackage{multicol}
\usepackage[linesnumbered,ruled]{algorithm2e}
\usepackage[accsupp]{axessibility}  % Improves PDF readability for those with disabilities.

\usepackage[pagebackref,breaklinks,colorlinks,citecolor=eccvblue]{hyperref}
\usepackage{orcidlink}

\begin{document}

\title{Phase-Guided Light Field for Spatial-Depth High Resolution 3D Imaging} 

\titlerunning{Phase-Guided Light Field for Spatial-Depth High Resolution 3D Imaging}

\author{Geyou Zhang\inst{1} \and
Ce Zhu\inst{1} \and
Kai Liu\inst{2} \and
Yipeng Liu\inst{1}}

\authorrunning{G. Zhang et al.}

\institute{University of Electronic Science and Technology of China \and
Sichuan University
}

\maketitle

\begin{abstract}
On 3D imaging, light field cameras typically are of single-shot, and however, they heavily suffer from low spatial resolution and depth accuracy. In this paper, by employing an optical projector to project a group of high-frequency phase-shifted sinusoid patterns, we propose a phase-guided light field algorithm to significantly improve both the spatial and depth resolutions for off-the-shelf light field cameras. First, in order to correct the axial aberrations caused by the main lens of our light field camera, we propose a deformed cone model to calibrate our structured light field system. Second, over wrapped phases computed from patterned images, we propose a stereo matching algorithm, i.e. phase-guided sum of absolute difference, to robustly obtain the correspondence for each pair of neighbored two lenslets. Finally, based on the reference depth by phase-guided stereo matching, we conduct a re-projection and refinement strategy to reconstruct 3D point clouds with spatial-depth high resolution. Experimental results show that, compared with the state-of-the-art active light field methods, the proposed reconstructs 3D point clouds with a spatial resolution of 1280$\times$720 with factors 10$\times$ increased, while maintaining the same high depth resolution and needing merely a single group of high-frequency patterns.
\end{abstract}
\section{Introduction}
\label{sec:intro}
  
Light field cameras captures image from multi-perspective arranged in an array form, thereby boasting a wide depth of field and abundant 3D scene geometry. Light field 3D imaging technologies can be categorized into two types: passive and active. Passive light field 3D imaging is of several noticeable advantages compared to binocular/multiocular 3D imaging, such as no need for hardware synchronization considerations and simpler system configuration. There are primarily four types~\cite{hog2017image} of passive light field 3D imaging methods: 1) sub-aperture image (SAI) based~\cite{tao2013depth,yu2013line,wang2016depth,srinivasan2017shape,park2017robust,jeon2018depth}, 2) lenslet image based~\cite{perwass2012single,johannsen2013calibration,heinze2016automated,heinze2015automated}, 3) epipolar plane image based~\cite{matouvsek2001accurate,wanner2011generating,wanner2012globally,zhang2016light}, and 4) refocused image based~\cite{kao2007depth,mousnier2015partial}. However, passive light field 3D imaging is not capable of high precision industrial inspection due to its limited accuracy.

Active light field 3D imaging, also as known as structured light field (SLF), projects encoded patterns onto the object to achieve high-accuracy 3D reconstruction. SLF is an emerging endeavor for 3D imaging having both the high accuracy of structured light illumination (SLI) and the implicit scene geometry information of light field. Based on the architecture of the commercial hand-held plenoptic camera 1.0~\cite{ng2005light}, various of researches have explored to significantly improve the accuracy of passive light field 3D imaging by employing active illumination. For SLF calibration, Cai et al proposed a ray calibration~\cite{cai2018ray} and a linear metric model~\cite{cai2018light} to relate depth values in object space and image space, but an extra auxiliary camera is needed. For 3D reconstruction, multiple-shot phase-shifting fringe patterns~\cite{cai2016structured,cai2018light,cai2020structured,Zhou22phase} are of high accuracy but excessive scanning time consumption, while single-shot sinusoid fringe pattern~\cite{cai2019accurate,cai2020single} and speckle pattern~\cite{wu2023dynamic} can achieve single-shot 3D imaging but sacrifice accuracy.

\begin{figure}
    \centering
    \includegraphics[width=1\linewidth]{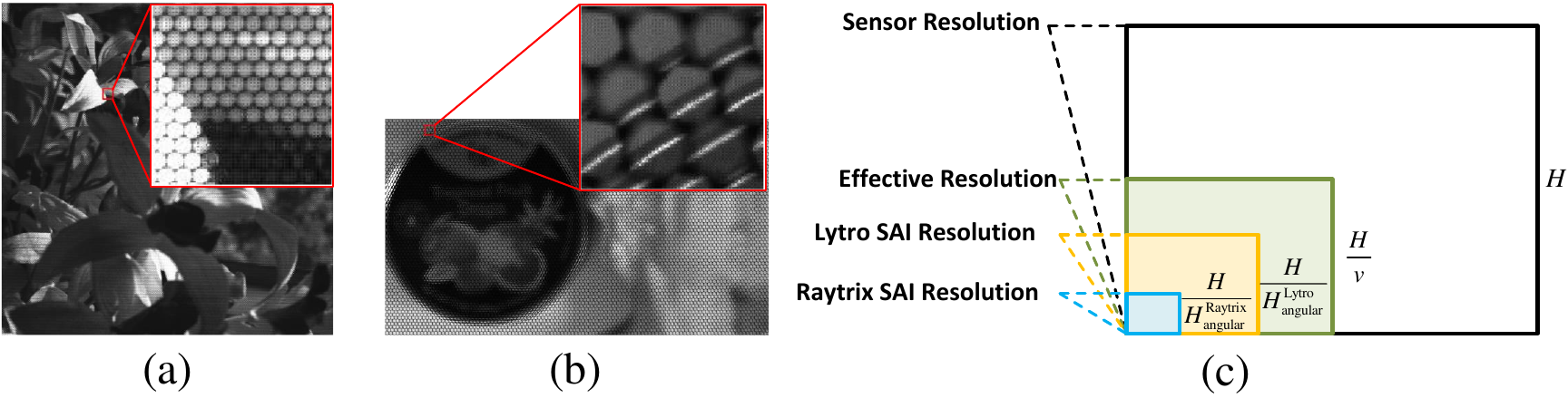}
    \caption{(a) Raw image of Lytro plenoptic camera 1.0, with $3280\times3280$ sensor resolution, $9\times9$ angular resolution, and $364\times364$ SAI resolution, (b) raw image of Raytrix plenoptic camera 2.0, with $3840\times2160$ sensor resolution, $35\times35$ angular resolution, and $110\times72$ SAI resolution and (c) effective resolution of light field imaging is associated with virtual depth $v$, which is determined by the number of lenslets that see the same target point along with the vertical or horizontal direction. The existing active light field technology can only obtain depth maps constrained by SAI resolution, far from accomplishing the effective resolution of a plenoptic camera.}
    \label{FIG:FigResolution} 
\end{figure} 

The existing active light field techniques employ plenoptic camera 1.0 with a low angular resolution of a few to a dozen pixels and a high SAI resolution (size of sub-aperture image) of hundreds of pixels, as shown in Fig.~\ref{FIG:FigResolution}(a), resulting in a considerable resolution of depth map~(filled with yellow in Fig.~\ref{FIG:FigResolution}(c)), but they failed to achieve the theoretical effective resolution~(filled with green in Fig.~\ref{FIG:FigResolution}(c)) of light field imaging. As a commercial-grade camera, the plenoptic camera 1.0 is not suitable for industrial conditions and has already stopped production. Different from plenoptic 1.0, in 2012, Raytrix company developed industrial-grade plenoptic camera 2.0 with an improved angular resolution to dozens of pixels, as shown in Fig.~\ref{FIG:FigResolution}(b), and currently has dominated the majority of the light field camera market share. The following problem is that, if the active light field algorithms tailored for plenoptic camera 1.0 are directly migrated to plenoptic camera 2.0, the loss of depth map resolution will become very serious due to the low SAI resolution (filled with blue in Fig.~\ref{FIG:FigResolution}(c)) of plenoptic camera 2.0.  

To sum up, we notice that the active light field techniques commonly suffer from the constraint of SAI resolution, preventing the depth resolution from achieving the effective resolution, this problem has become more severe in the current plenoptic camera 2.0. Consequently, the application of active light field technique in precision industrial inspection is hindered by the loss of depth map spatial resolution. Consequently, a natural question emerges: \textbf{How} can we eliminate the resolution constraint on the depth map in active light field 3D imaging?

\begin{figure}
    \centering
        \includegraphics[width=0.958\textwidth]{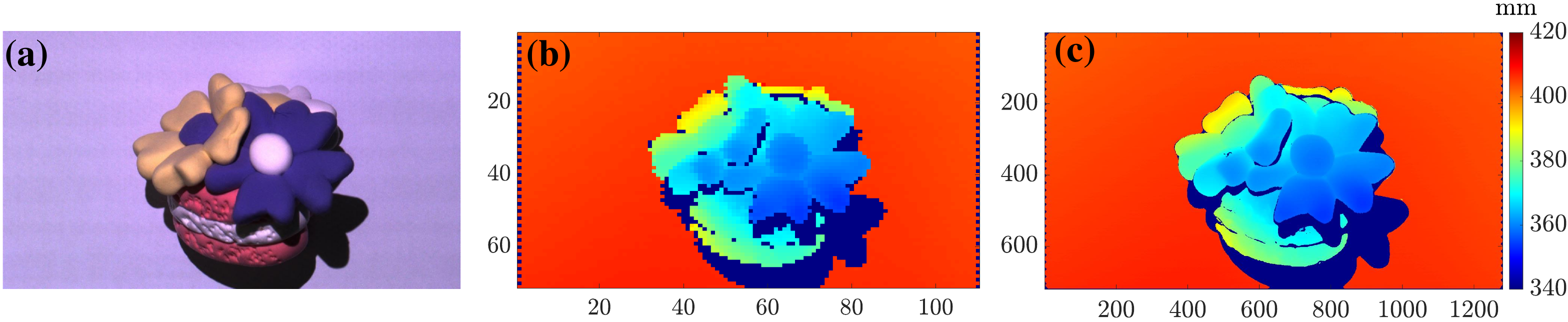}
      \setlength{\linewidth}{\textwidth}
      \setlength{\hsize}{\textwidth}
    \caption{(a) Measured object, (b) state-of-the-art active light fields~\cite{cai2016structured,cai2018light,cai2020structured} employ SAI array resulting in low spatial resolution depth map (resolution: 110$\times$71), and (c) our PGLF works on lenslet image array to have high spatial resolution depth map (resolution: 1280$\times$720).} 
    \label{FIG:QualitativeEvaluation}
\end{figure}

To address this, in this paper, we present a phase-guided light field (PGLF) 3D imaging pipeline for plenoptic camera 2.0 architecture, the spatial resolution of the depth map is improved by a factor of 10$\times$ compared with state-of-the-art active light field techniques as shown in Fig.~\ref{FIG:QualitativeEvaluation}. Moreover, by leveraging the implicit 3D scene geometry of light field images, the high-frequency ambiguity of the wrapped phase can be eliminated, thereby achieving unambiguous 3D reconstruction from only a single group of high-frequency patterns, thus significantly reducing scanning time consumption.

\section{Contributions} 
In the existing passive and active light field techniques, the depth map resolution is limited by the SAI resolution. Our method breaks this limitation and improves the depth map resolution up to the optical effective resolution~\cite{perwass2012single} of the plenoptic camera. Our method is tailored for plenoptic camera 2.0 with high angular resolution, for instance, consider a Raytrix R8 plenoptic camera with an angular resolution of $35\times35$ and spatial resolution of $110\times71$, i.e., a sensor resolution of $3840\times2160$, on the measurement circumstance with a virtual depth of approximately 3, if we ignore the defocus of the micro-lens, theoretically, our method accomplishes an effective resolution ratio~\cite{perwass2012single}~(ERR, the ratio of effective resolution and sensor resolution) of $1/3$ and achieves 3D imaging with a resolution of $1280\times720$, while previous SLF methods is of the ERR of $1/35$ and can merely achieve 3D imaging with a resolution of $110\times71$.  
    
To the best of our knowledge, we are the first to relieve the limitation on depth map resolution constrained by SAI resolution in active light field imaging. Our method enables the plenoptic camera 2.0 to approach its optical limitation for 3D imaging, resulting in a 10$\times$ enhancement in the resolution of depth images when compared to existing active and passive light field 3D imaging techniques. This improvement empowers light field 3D imaging for dense and precise 3D measurements, such as defect detection and reverse engineering. Moreover, by exploiting the reference depth map generated by our PGLF algorithm, only one group of fringe is needed for 3D reconstruction, thereby significantly reducing time consumption by half. In applications such as automatic optical inspection, this efficiency gain translates into considerable time savings.

    \begin{figure}[!h]
		\centering
		\includegraphics[width=0.9\linewidth]{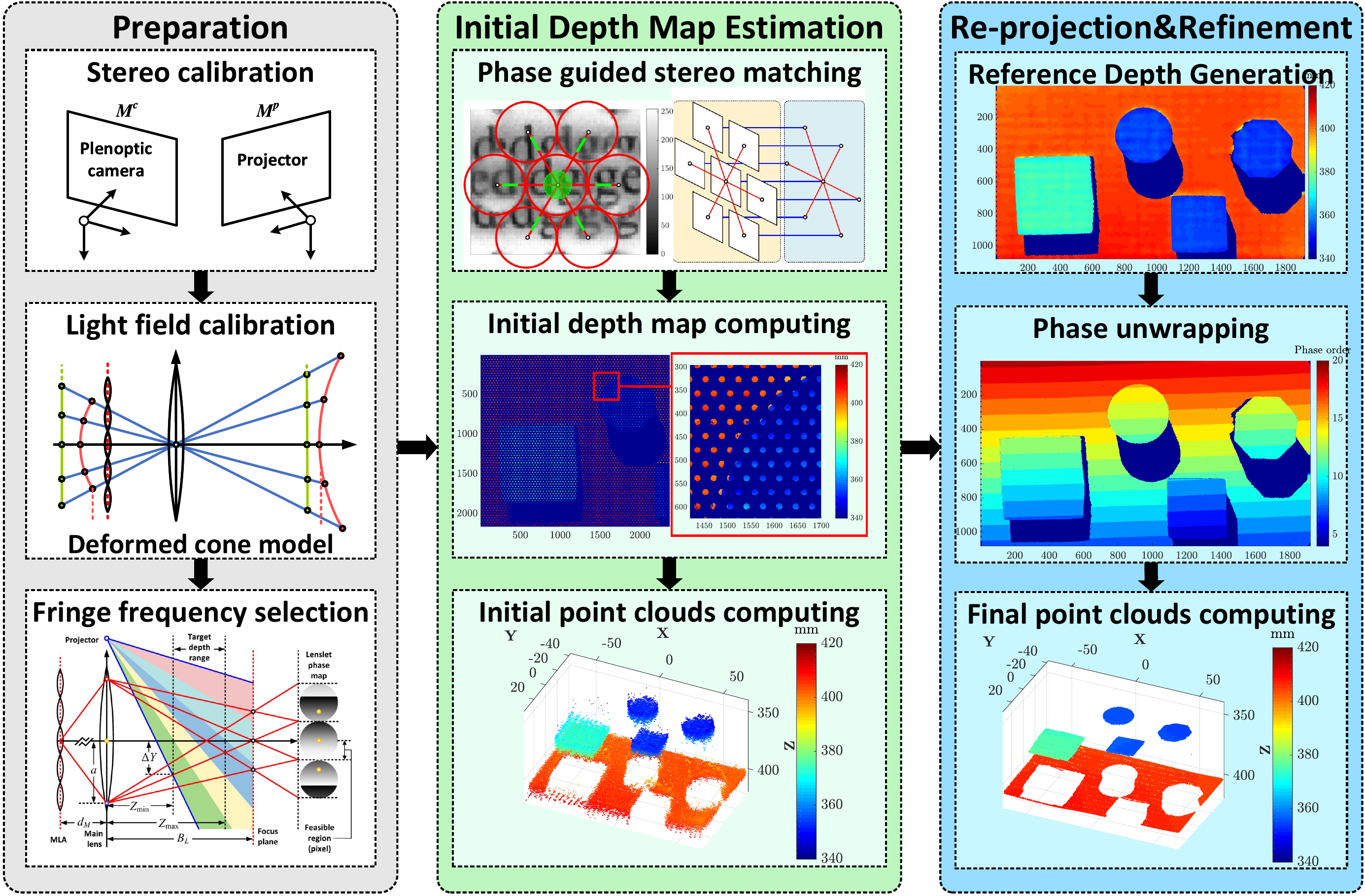}
		\caption{Flowchart of our method (gray box: calibrate structure light field system with our DCM, green box: initial depth map by phase-guided light field, and blue box: re-projection and refinement strategy for dual-high resolution).}
		\label{Fig:FigFlowchart}
    \end{figure}
 
	The procedure of our method is shown in Fig.~\ref{Fig:FigFlowchart}, which clarifies that the main contributions of this paper are:
	
	\hangindent=0.8cm
        $\bullet$~SLF system calibration procedure with a deformed cone model (DCM) to address axial aberration issue.
    
	\hangindent=0.8cm
	$\bullet$~Phase-guided sum of absolute difference (PSAD) cost to robustly achieve stereo matching between adjacent lenslet images.
	
	\hangindent=0.8cm
	$\bullet$~Re-projection and refinement strategy for generating spatial-depth high resolution 3D point clouds.

\section{Preparation for SLF 3D Imaging}
    Passive light field is challenged by weak texture and depth discontinuity issues. To tackle these problems, we adopt an active stereo strategy and project at least 3 phase-shifting fringe patterns as $I_n^p\left( x^p, y^p \right) = \frac{1}{2} + \frac{1}{2}\cos \left( \frac{2 \pi fy^p}{H^p}- \frac{2 \pi n}{N}\right)$, where, $( x^p, y^p )$  is the coordinate in the projector space, the term $I_n^p$ is the grayscale intensity of the pixel, the integer-valued $f$ is the spatial frequency of the fringes, $H^p$ is the height of the projector's spatial resolution in rows, and $n$ and $N$ are the index and total number of the phase shifts. Then the wrapped phase can be computed as
    
	\begin{equation}
		\phi = \tan ^{-1}\left[\frac{\sum_{n = 0}^{N - 1} I_n^c \sin \left(\frac{2\pi n}{N}\right)} {\sum_{n = 0}^{N - 1} I_n^c \cos \left(\frac{2 \pi n}{N}\right)} \right],
		\label{EQ:EqComputePhase}
	\end{equation} 
	
    Our SLF system consists of a plenoptic camera and a projector. Two calibrations need to be performed to comprehensively describe the system: 1) stereo calibration: calibrate the camera-projector pair; 2) light field calibration: calibrate plenoptic camera parameters.
	
    \subsection{Stereo Calibration}
    We create a virtual camera overlapping the position of the plenoptic camera, the resolution is set to the effective resolution of the plenoptic camera under the actual virtual depth. For example, when we use a plenoptic camera that boasts a sensor resolution of 3840$\times$2160 to measure objects with a virtual depth of approximately 3, the virtual camera resolution should be correspondingly set to 1280$\times$720. The images captured by the virtual camera are rendered from light field images by our phase-guided stereo matching (Section 4.1). Our stereo calibration is for the virtual camera and projector. 

    The camera-projector pair can be described by perspective projection matrices $\textbf{M}^c$ and $\textbf{M}^p$, which can be solved by least squared method~\cite{yalla2005very}. Once the camera-projector pair is calibrated, the triangulation based 3D reconstruction typically computes point clouds~\cite{liu2010dual} by
	\begin{equation}\label{EQ:Conventional3DReconst}
		{\left[ {{X}\quad {Y}\quad {Z}} \right]^{\text{T}}} = G_0^{ - 1}h,
	\end{equation}
    where $G$ and $h$ are derived from $\textbf{M}^c$ and $\textbf{M}^p$. Note that the world coordinate is established to coincide with the camera coordinate system in our calibration.
 
    \subsection{Light field Calibration}
    An ideal plenoptic camera follows a linear imaging model~\cite{heinze2015automated}	as
	\begin{equation}\label{EQ:LightFieldDepth}
			Z = \frac{z f_L}{z - f_L},z = v d_\mu + d, v = \frac{D_\mu }{D_\mu - D}\\
	\end{equation}
    where $z$ is the image distance of the target 3D point $(X, Y, Z)$, $v$ is the virtual depth, $D$ is the corresponding point distance in the adjacent lenslet images, moreover, prior optical system parameters: $d_\mu$ is the distance between the sensor plane and the MLA plane, $D_\mu$ is the diameter of each lenslet in pixels, $f_L$ is the focal length of the main lens. 	
 
 	\begin{figure}[h!] 
		\centering
            \begin{minipage}[c]{0.5\linewidth} 
            \subfloat[]
		{
    		\includegraphics[width=\linewidth]{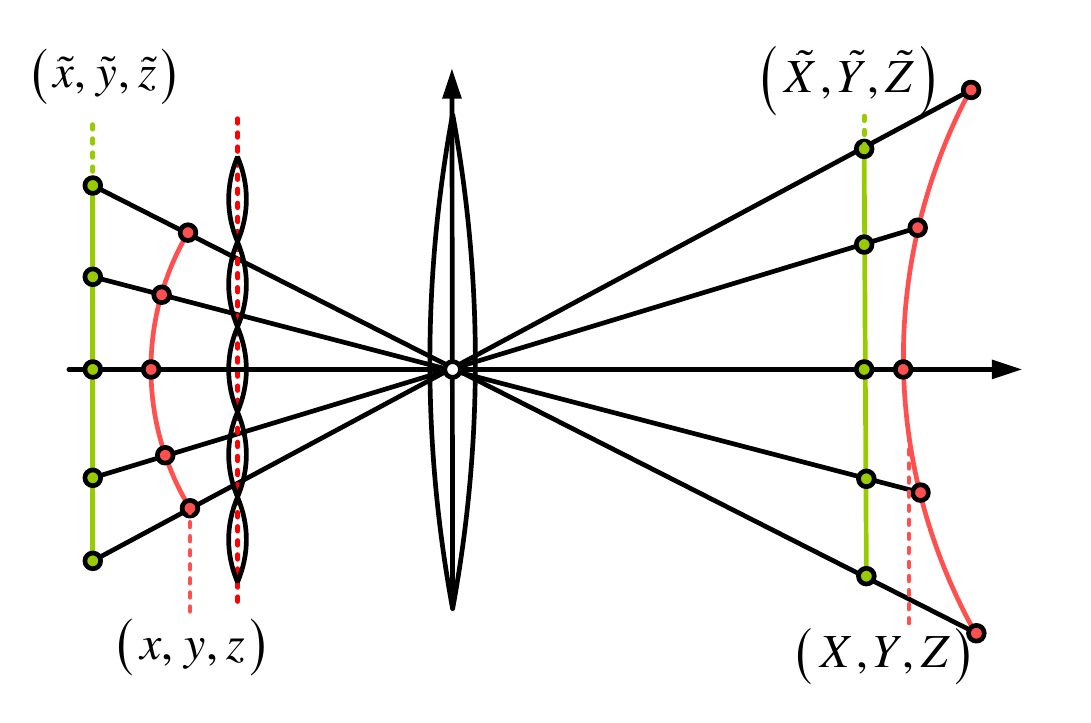}
		}
            \end{minipage}
          \begin{minipage}[c]{0.24\linewidth} 
              	\subfloat[]
        		{
        			\includegraphics[width=\linewidth]{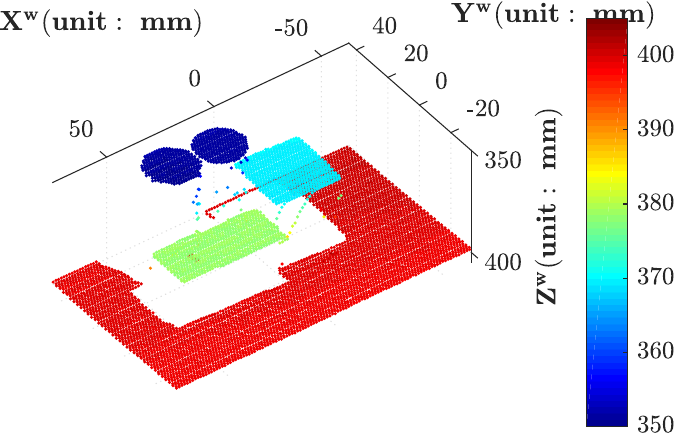}
        		}\\
          		\subfloat[]
        		{
        			\includegraphics[width=\linewidth]{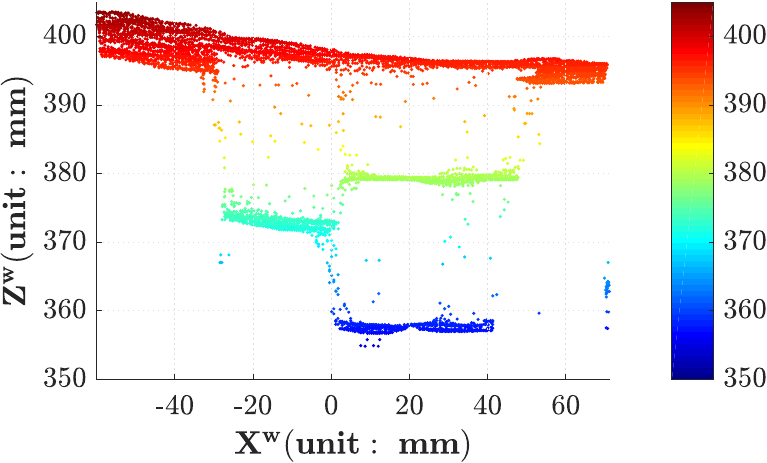}
        		}
          \end{minipage}
          \begin{minipage}[c]{0.24\linewidth} 
        		\subfloat[]
        		{
        			\includegraphics[width=\linewidth]{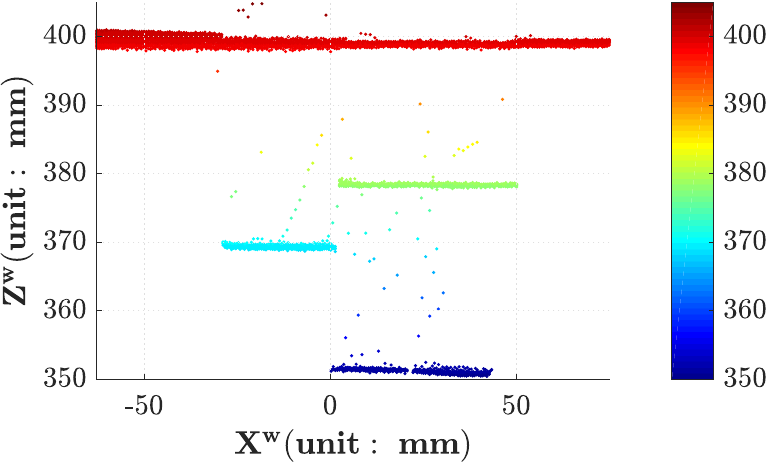}
        		}\\
        		\subfloat[]
        		{
        			\includegraphics[width=\linewidth]{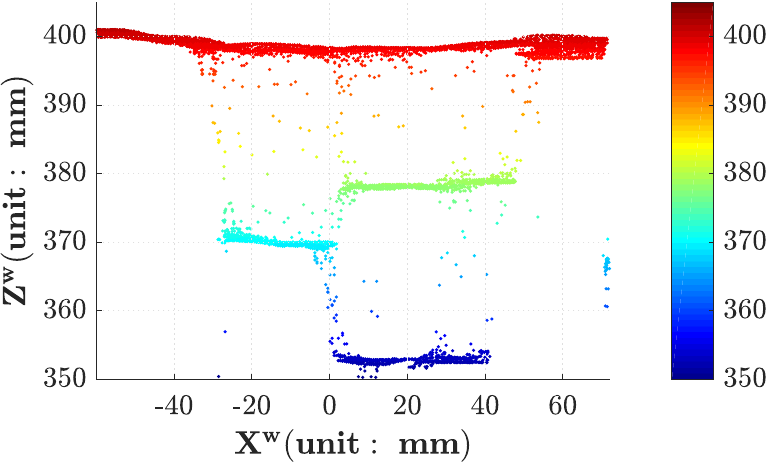}
        		}
          \end{minipage}

		\caption{ (a) Our DCM for describing axial aberration.    Point clouds of the gypsum blocks: (b), (d) ground truth computed by Eq.~(\ref{EQ:Conventional3DReconst}), (c) linear model (Eq.~(\ref{EQ:Conventional3DReconst})), and (e) our DCM.} 
		\label{FIG:FigDepthWrapping}
	\end{figure}
 
    However, the actual optical imaging process suffers from aberrations in the main lens, which causes the depth deformation phenomenon and significantly reduces the accuracy of light field 3D imaging as exemplified in Fig.~\ref{FIG:FigDepthWrapping}(c). This is primarily attributed to axial aberration: field curvature and astigmatism. Therefore, as demonstrated in Fig.~\ref{FIG:FigDepthWrapping}(a), assuming that the field curvature and astigmatism are functions of the incident angle of the chief ray $(\theta_x,\theta_y)$ and the image distance $z$ (equivalent to virtual depth~\cite{perwass2012single} $v$), we propose a DCM for characterizing the imaging distance $\tilde{z}$, free from axial aberrations, as a function of $(\theta_x,\theta_y, v)$ with the aid of parameters $\textbf{a}$, $k$, and $d$:
    \begin{equation}\label{EQ:AxialAberrationModel}
		{\tilde{z}}\left( {\theta_x,\theta_y,v} \right) = \textbf{a}^{\text{T}} \textbf{p} \left( 1 + k v \right)\left( vd_{\mu} + d \right),
    \end{equation}
    where
    \begin{equation}\nonumber
    \setlength{\arraycolsep}{5pt}
		\left\{ \begin{split}
			\textbf{a} &= {\left[ {\begin{array}{*{40}{c}}
						1&{{a_1}}&{{a_2}}&{{a_3}}&{{a_4}}&{{a_5}}& 
				\end{array}} \right]^{\text{T}}}\\
			\textbf{p} &= {\left[ {\begin{array}{*{40}{c}}
						1&{\theta_x}&{\theta_y}&{{\theta_x}{\theta_y}}&{\theta_x^2}&{\theta_y^2} 
				\end{array}} \right]}
		\end{split}  \right. 
    \end{equation}
	and
    \begin{equation}\label{EQ:IncidentAngle}
		\left[ \begin{gathered}
			{\theta _x} \\
			{\theta _y} \\ 
		\end{gathered}  \right] = \frac{1}{{{Z}}}\left[ \begin{gathered}
			{X} \\
			{Y} \\ 
		\end{gathered}  \right] = \frac{1}{{{z}}}\left[ \begin{gathered}
			{x} \\
			{y} \\ 
		\end{gathered}  \right].
    \end{equation}

	To calibrate the parameter $\textbf{a}$, we successively scan a plate for $M$ times within the desired depth range. Then we select $T$ lenslet center pixels as the target points. Therefore, we can numerically solve $\textbf{a}$ by nonlinear least squared method as
	\begin{equation}\label{EQ:dMinimizing}
		\left[\textbf{a},k,d\right] = \mathop {\arg \min }\limits_{\left[\textbf{a},k,d\right]} {\left\| \textbf{\textit{z}} - {\tilde {\textbf{\textit{z}}}\left( {\theta_{x,m,t},\theta_{y,m,t},{v_{m,t}}} \right)} \right\|},
	\end{equation}
    where $\|\cdot\|$ denotes Euclidean norm, $\textbf{\textit{z}}$ and $\tilde{\textbf{\textit{z}}}$ are vectors computed with $ z_{m,t} = Z_{m,t} f_L/(Z_{m,t} - f_L)$ and $\tilde{z}_{m,t} = \tilde{z}(\theta_{x,m,t},\theta_{y,m,t},v_{m,t})$ during the $M$ times scanning with $T$ target points each time, respectively, in which $Z_{m,t}$ and $v_{m,t}$ are computed by Eq.~(\ref{EQ:Conventional3DReconst}) and phase-guided stereo matching (we will illustrate in section 4.1), respectively. 
	
    By replacing $z$ in Eq.~(\ref{EQ:LightFieldDepth}) with $\tilde{z}$, we can effectively compensate the depth warping induced by axial aberration in light field 3D imaging as shown in Fig.~\ref{FIG:FigDepthWrapping}(d).   
 
\subsection{Calibration Procedure}
	\begin{algorithm}
		\small
		% \setstretch{1.2}
		\caption{SLF system calibration} 
		\label{ALG:AlgCalibration}
		\KwIn
		{							
			\\Captured images of LF calibration plates: $\left\{P_{m,n}\right\}$.
			\\Captured images of stereo calibration target: $\left\{Q_n\right\}$.	
		}
		\KwOut
		{
			\\Virtual camera and projector matrices: $\textbf{M}^c$, $\textbf{M}^p$.	
			\\ Plenoptic camera intrinsics: ${\textbf{a},k,d}$.		
		}
		% \tcp{Stereo Calibration}
		$\phi^Q\leftarrow$ \text{ComputePhase}$(\{Q_n\})$
		
		$D^Q\leftarrow$ \text{LensletPhaseMatching}$(\phi^Q)$
		
		$\{Q_n^r\}\leftarrow$ \text{Refocus}$(\left\{Q_n\right\},D^Q)$
		
		$\textbf{M}^c,\textbf{M}^p\leftarrow$ \text{StereoCalibration}$(\{Q_n^r\})$

		% \tcp{Light Field Calibration}
		\For{ $m \leftarrow 1$ \KwTo $M$ }	
		{		
			$\phi^P_m\leftarrow$ \text{ComputePhase}$(\{P_{m,n}\})$
			
			$(\textbf{\textit{X}}_m,\textbf{\textit{Y}}_m,\textbf{\textit{Z}}_m)\leftarrow$ \text{Compute3DSLI}$(\phi^P_m,\textbf{M}^c,\textbf{M}^p)$
			
			$\textbf{\textit{D}}^P_m\leftarrow$ \text{LensletPhaseMatching}$(\phi^P_m)$
			
			$\tilde{\textbf{\textit{z}}}_m \leftarrow \tilde{z}(\frac{\textbf{\textit{X}}_m}{\textbf{\textit{Z}}_m},\frac{\textbf{\textit{Y}}_m}{\textbf{\textit{Z}}_m},\frac{D_\mu }{D_\mu - \textbf{\textit{D}}^P_m})$
			
			$\textbf{\textit{z}}_m \leftarrow \frac{\textbf{\textit{Z}}_m f_L}{\textbf{\textit{Z}}_m - f_L}$		
   
		}
		$\textbf{\textit{z}} \leftarrow$ \text{Concat} $\textbf{\textit{z}}_1,...,\textbf{\textit{z}}_M$
		
		$\tilde{\textbf{\textit{z}}} \leftarrow$ \text{Concat} $\tilde{\textbf{\textit{z}}}_1,...,\tilde{\textbf{\textit{z}}}_M$
		
		$\left[\textbf{a},k,d\right] \leftarrow \mathop {\arg \min }\limits_{\left[\textbf{a},k,d\right]} {\left\| {\textbf{\textit{z}} - \tilde{\textbf{\textit{z}}}} \right\|}$
	\end{algorithm}
 
    Note that the phase in the feasible region should be single-valued, which is an indispensable guarantee for stereo matching. Thus, we propose a \textbf{uniqueness constraint} for fringe frequency selection. Implementation details can be found in the supplementary material. Our proposed calibration method for SLF system is summarized by Algorithm~\ref{ALG:AlgCalibration}. We list the notations in Algorithm~\ref{ALG:AlgCalibration}: ComputePhase: Eq.~(\ref{EQ:EqComputePhase}), LensletPhaseMatching: Section 4.1, Refocus: Section 5.3, Compute3DSLI: Eq.~(\ref{EQ:Conventional3DReconst}), StereoCalibration: Ref.~\cite{yalla2005very}, and Concat: Concatenate elements.

\section{Initial Depth Map Estimation}
    Phase can be treated as a robust texture for stereo matching of adjacent lenslet images. Before we conduct the phase-guided stereo matching, to avoid invalid calculations, the feasible region of corresponding point distance $\left[ D_\mathrm{min},D_\mathrm{max} \right]$ and the radius $r$ of the circularly-shaped area to be computed in each lenslet image can be determined according to the desired depth range $\left[ Z_\mathrm{min},Z_\mathrm{max} \right]$, as shown in Fig.~\ref{FIG:FigPSADWeight}(b).

    \subsection{Phase-Guided Stereo Matching on Lenslets}	
 
\begin{figure}[!h]
		\centering
            \begin{minipage}[c]{0.3\linewidth} 
            \subfloat[]
            {
               \centering
               \includegraphics[width=\linewidth]{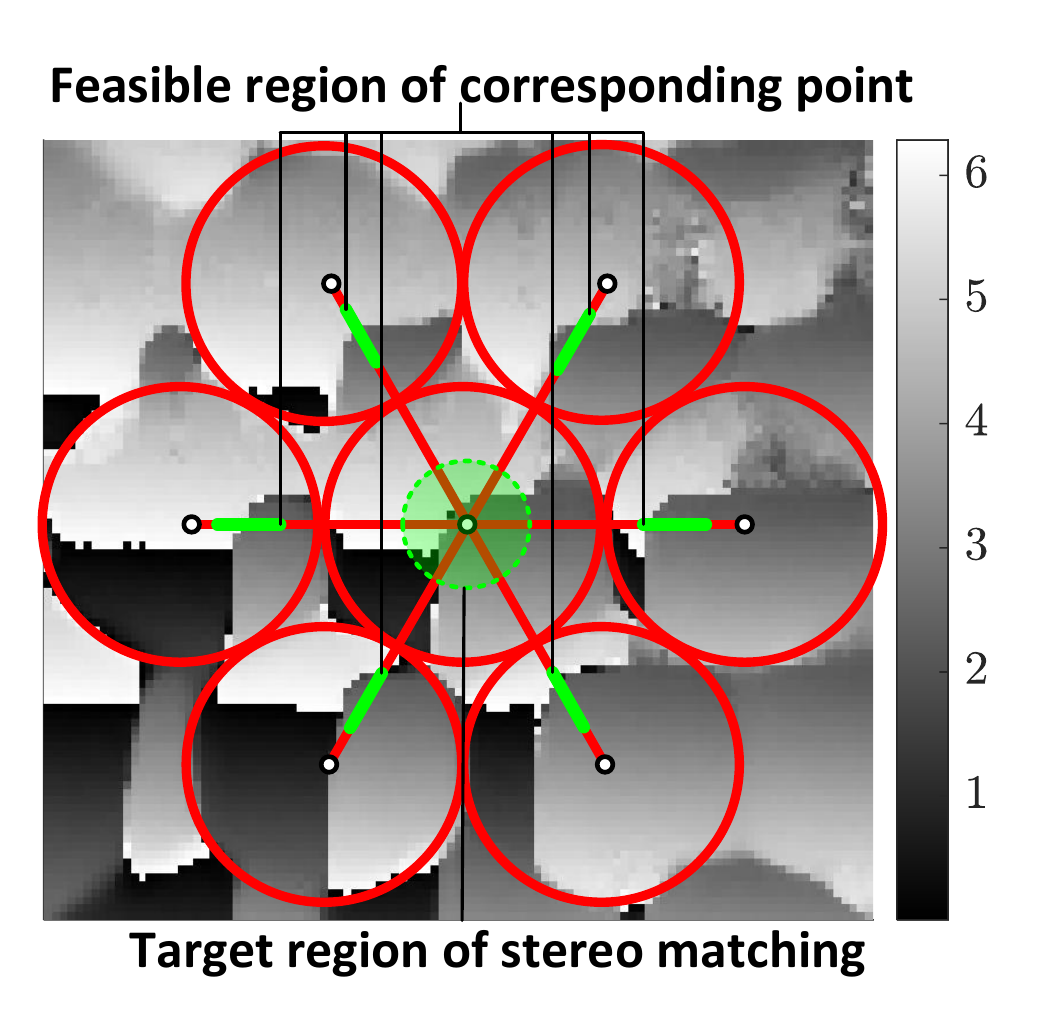}		
            }
            \end{minipage}
            \begin{minipage}[c]{0.68\linewidth}                 
		\subfloat[]
		{
			\centering
			\includegraphics[width=0.2967\linewidth]{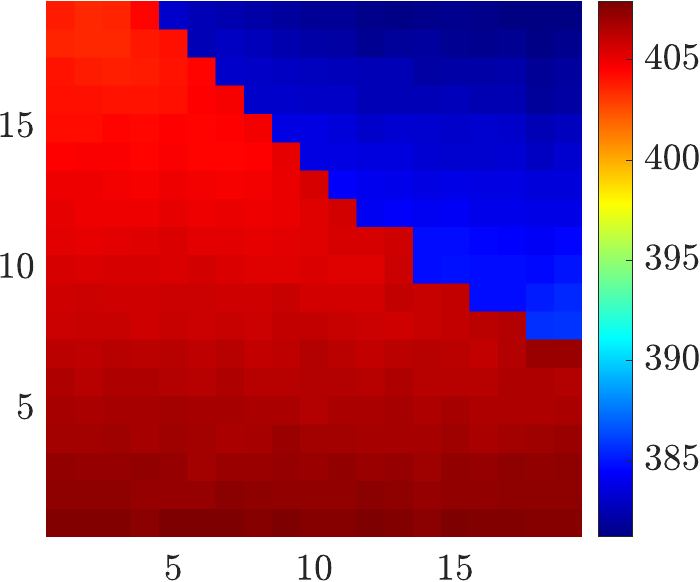}
		}
		\subfloat[]
		{
			\centering
			\includegraphics[width=0.4066\linewidth]{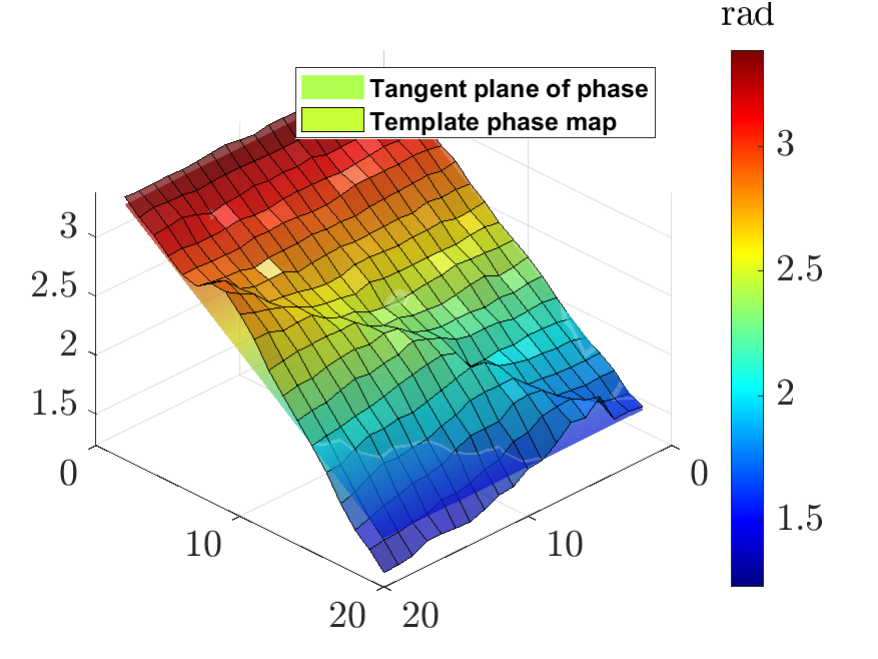}
		}
		\subfloat[]
		{
			\centering
			\includegraphics[width=0.2967\linewidth]{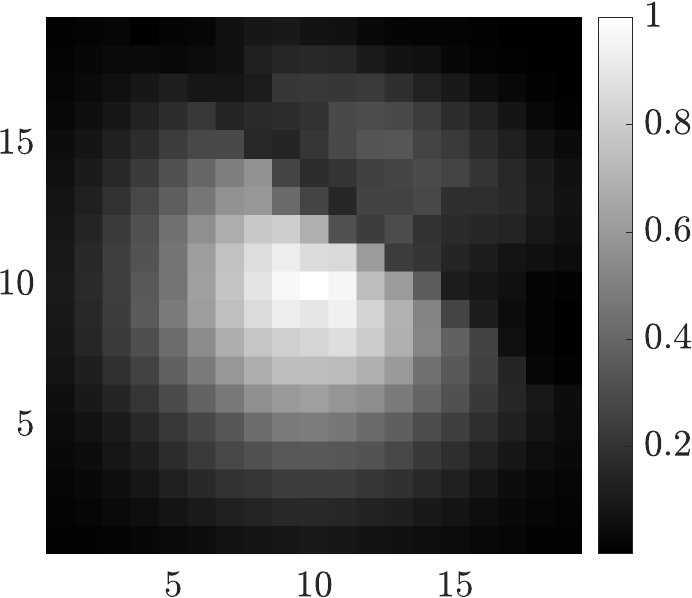}
		}
		\\
		\subfloat[]
		{
			\centering
			\includegraphics[width=0.2967\linewidth]{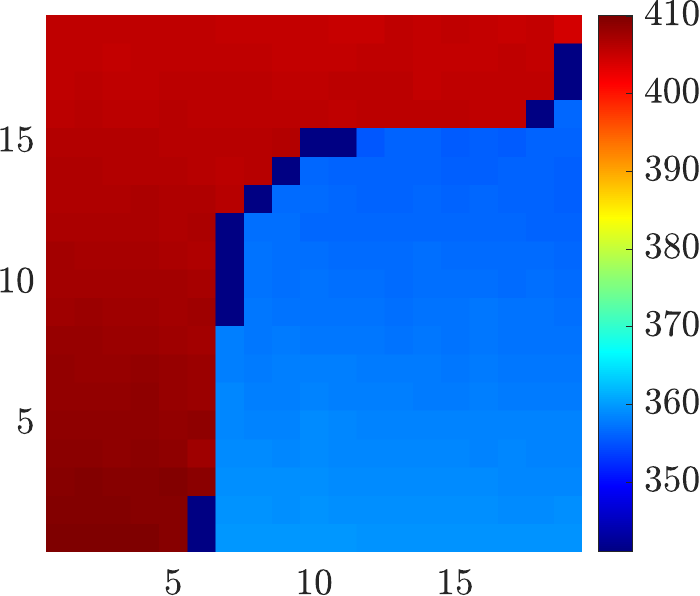}
		}
		\subfloat[]
		{
			\centering
			\includegraphics[width=0.4066\linewidth]{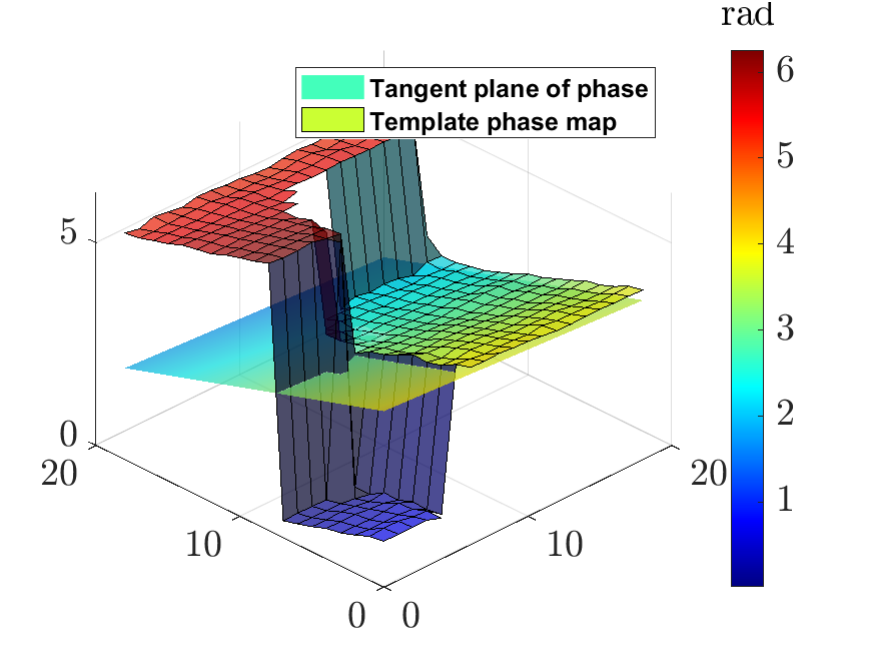}
		}
		\subfloat[]
		{
			\centering
			\includegraphics[width=0.2967\linewidth]{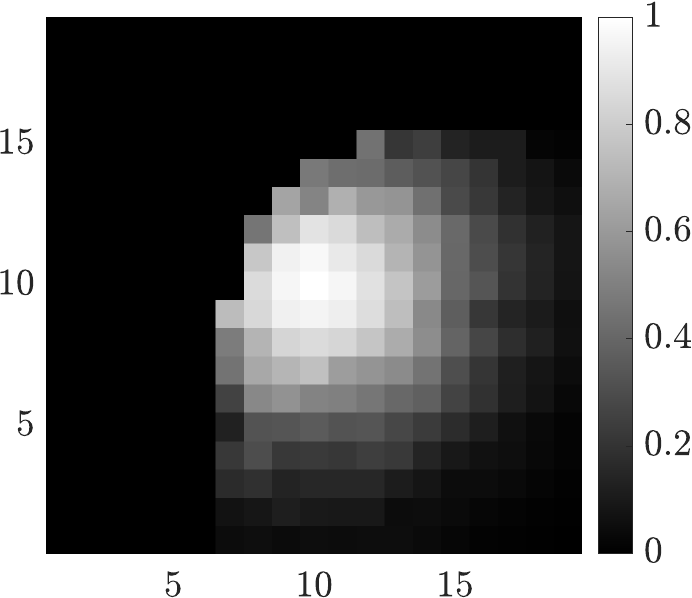}
		}
            \end{minipage}
		\caption{(a) Diagram of phase-guided stereo matching (phase map of (e), (f), and (g)). (b) and (e) depth, (c) and (f) $\phi$ and $\tilde{\phi}$, (d) and (g ) $W_0$, within two lenslet images.} 
		\label{FIG:FigPSADWeight}
	\end{figure}

	We select six lenslet images circled with the red solid line in Fig.~\ref{FIG:FigPSADWeight}(a) for phase-guided stereo matching. Due to the directionality of the phase, we exclude the target lenslets that hold $\frac{\textbf{g} \cdot \textbf{u}}{|\textbf{g}||\textbf{u}|} < \cos\left(\pi/3\right)$, where the phase gradient $\textbf{g}=\left[ G_x~~G_y \right]$, $G_x$ and $G_y$ are the medians of the horizontal and vertical gradient of $\phi$ within a $7\times7$ square window centered at the template pixel, $\textbf{u}$ is the unit vector from the template to target lenslet, from which we ensure the phase gradient maintains a sufficient component along the epipolar. 
 
	Different from the stereo matching of grayscale images, we focus on the absolute value rather than the gradient or relative value in phase-based matching. Thus, the sum of absolute differences (SAD) is relatively appropriate among the block-matching methods. However, its accuracy diminishes when the occluder in the lenslet image begins to dominate over non-occluded pixels, resulting in errors at depth discontinuities. 
 
	For suppressing the depth discontinuous error, we propose a PSAD algorithm for stereo matching, based on two key assumptions to increase the robustness,  1) \textbf{phase adjacency}: in the wrapped phase $\phi$, the closer the phase difference between two pixels is to 0 or $2\pi$, the more likely they are to be adjacent in 3D space, and 2) \textbf{pixel adjacency}: the closer the distance between two pixels, the more likely they are to be adjacent in 3D space. Consequently, spatial adjacency can be used for weighting the cost function. Moreover, to more precisely describe the spatial adjacency, the phase gradient should be considered. Thus, we model the tangent plane of phase with the gradient information of phase map neighboring the template pixel $\textbf{\textit{s}}^c_j$ as
	\begin{equation}
		{{\tilde \phi }}\left( \textbf{\textit{s}}^c \right) = {\phi}\left( \textbf{\textit{s}}^c_j \right) + \emph{\textbf{g}}_j 	{\left(\textbf{\textit{s}}^c - \textbf{\textit{s}}^c_j \right)^{\text{T}}},
	\end{equation}
    where $\textbf{\textit{s}}^c=(x^c,y^c)$ denotes a arbitary 2D index of the phase map $\phi$.
 
    Further, the weight of block matching can be given as
    \begin{equation}\label{EQ:EqWeight}
			W\left( \textbf{\textit{s}}^c \right) = \left\{ \begin{gathered}
				1,W_0\left( \textbf{\textit{s}}^c \right) \geq \tau_1 \\
				0,W_0\left( \textbf{\textit{s}}^c \right) < \tau_1
			\end{gathered}  \right.,
    \end{equation}
        where 
    \begin{equation}
    \nonumber        			
    \left\{ \begin{aligned}
                W_0\left( \textbf{\textit{s}}^c \right) &= \exp \left( { - \frac{\|\textbf{\textit{s}}^c- \textbf{\textit{s}}^c_j\|}{{2{\sigma_s^2}}}} \right) \exp \left( { - \frac{{\Delta \phi^2\left( \textbf{\textit{s}}^c \right)}}{{2{\sigma_\phi ^2}}}} \right)\\
                \Delta\phi\left( \textbf{\textit{s}}^c \right) &= \min \left(\Delta\phi_0\left( \textbf{\textit{s}}^c \right),\left|\Delta\phi_0\left( \textbf{\textit{s}}^c \right) - 2\pi \right|\right) \\
                \Delta\phi_0\left( \textbf{\textit{s}}^c \right) &= \left|{{\tilde{\phi}}\left( \textbf{\textit{s}}^c \right) - {\phi}\left( {\textbf{\textit{s}}^c}\right)}\right|
			\end{aligned} \right.,
   \end{equation}
    truncation threshold $\tau_1$ is set to 0.4, $\sigma_s$ is set to 0.5$w$, the square window width $w$ is set to 13, and $\sigma_\phi$ is set to $3|\textbf{g}\textbf{u}|$  for adapting to various regions with different phase gradients. The homogeneous pixels of $\textbf{\textit{s}}^c_j$ are assigned higher $W_0$ as depicted in Fig.~\ref{FIG:FigPSADWeight}.               
    Finally, the PSAD cost is
	\begin{equation}\label{EQ:PSADCost}		
		C\left( \textbf{\textit{s}}^c_j, D \right) =  \sum\limits_{\textbf{\textit{s}}^c \in \mathcal{E} } {\frac{\min \left( {W\left( \textbf{\textit{s}}^c \right)\Delta {\phi}\left( \textbf{\textit{s}}^c,D \right),\tau_2 } \right)}{N_\mathcal{E}}},
	\end{equation}
	where
	\begin{equation}
        \nonumber
        \label{EQ:PSADCost2}
		\left\{
             \begin{aligned}
			\Delta {\phi}\left( \textbf{\textit{s}}^c, D \right) &= \min \left(\Delta\phi_0\left( \textbf{\textit{s}}^c,D \right),\left|\Delta\phi_0\left( \textbf{\textit{s}}^c,D \right) - 2\pi \right|\right) \\
			\Delta\phi_0\left( \textbf{\textit{s}}^c,D \right) &= \left| {{\phi}\left( \textbf{\textit{s}}^c \right) - {\phi}\left({\textbf{\textit{s}}^c + D\textbf{u}} \right)} \right|
            \end{aligned}\right.,
	\end{equation}
	 $\tau_2$ is set to $2|\textbf{g}\textbf{u}|$, $\mathcal{E}$ is a set of the valid adjacent pixels within a $w\times w$ rectangular window centered at $\textbf{\textit{s}}^c_j$, $N_\mathcal{E}$ is the number of elements in $\mathcal{E}$, and $D \in [D_{\mathrm{min}}, D_{\mathrm{max}} ]$. By minimizing $C\left( \textbf{\textit{s}}^c_j, D \right)$, we can compute the corresponding point distance $D$. As a result, The depth discontinuous error is effectively suppressed in the stereo matching results (i.e., reference depth map, we will illustrate how to generate it in the subsequent section) as shown in Fig.~\ref{FIG:FigDepth}(e)-(g). 	 	

    After stereo matching, we average the valid corresponding point distances to have $D$, and put $D$ into Eq.~(\ref{EQ:LightFieldDepth}), the initial depth $Z_r$ can be computed and shown in Fig.~\ref{FIG:FigDepth}(a), which visually presents a regularly distributed speckle pattern. 
 
    \subsection{Initial 3D Point Clouds Computing}		
	$X_r$ and $Y_r$ are needed to correct the depth deformation with incident angle $(\theta_x, \theta_y)$. According to the geometrical optics in a plenoptic camera, we derive the expression as
	\begin{equation}\label{EQ:XwAndYw}
		\left[ \begin{gathered}		    
		{X_r} \\{Y_r} \end{gathered}\right] = \emph{\textbf{k}}{Z_r} + \emph{\textbf{b}}.
	\end{equation}
    Implementation details can be found in the supplementary material. Then we compute the incident angle $(\theta_x, \theta_y)$ by Eq.~(\ref{EQ:IncidentAngle}), and subsequently substitute $(\theta_x, \theta_y, v)$ into Eq.~(\ref{EQ:AxialAberrationModel}) to have corrected depth $\tilde{Z}_r$. Finally, we compute $\tilde{X}_r$ and $\tilde{Y}_r$ with $(\theta_x, \theta_y, \tilde{Z}_r)$ to have initial 3D point clouds.

\section{Re-projection and Refinement Strategy for Accurate Point Clouds}
 	\begin{figure}[!h]
		\centering
                \subfloat[]
            {
		      \includegraphics[width=0.45\linewidth]{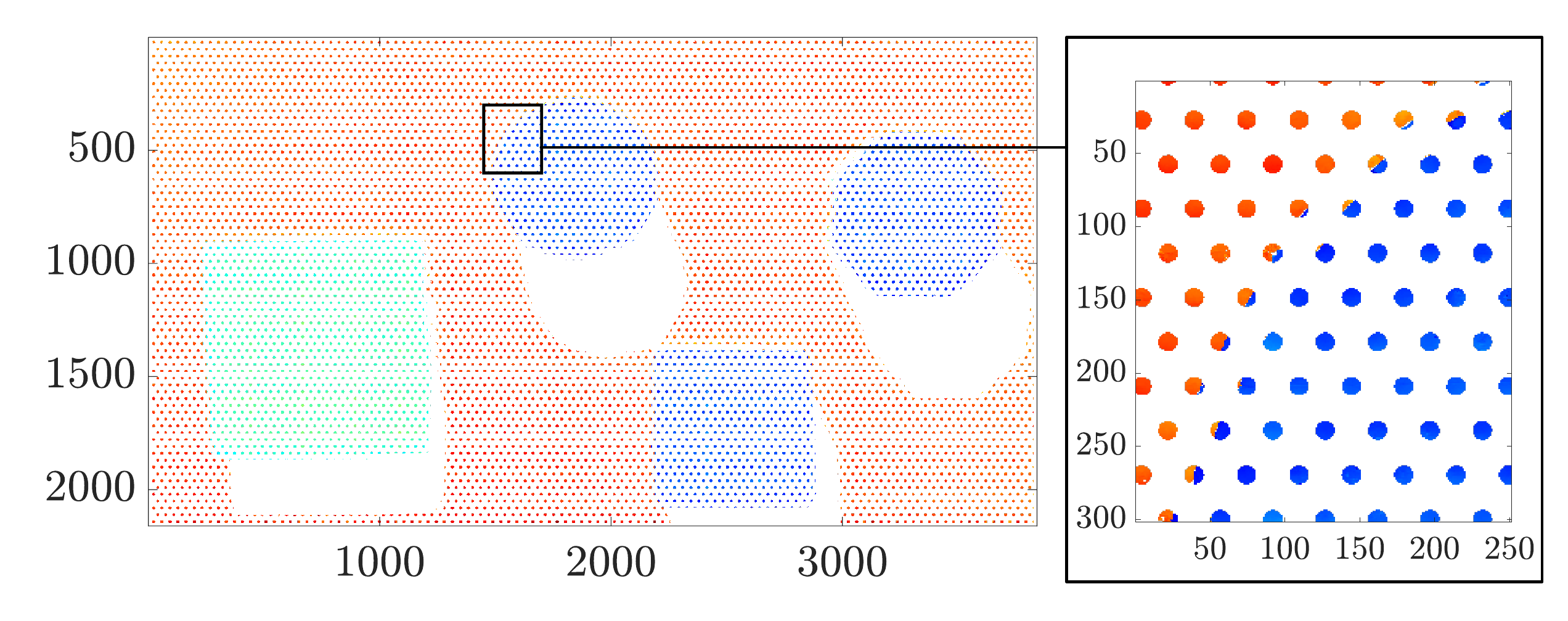}
              }
              \subfloat[]
            {
                \centering
                \includegraphics[width=0.45\linewidth]{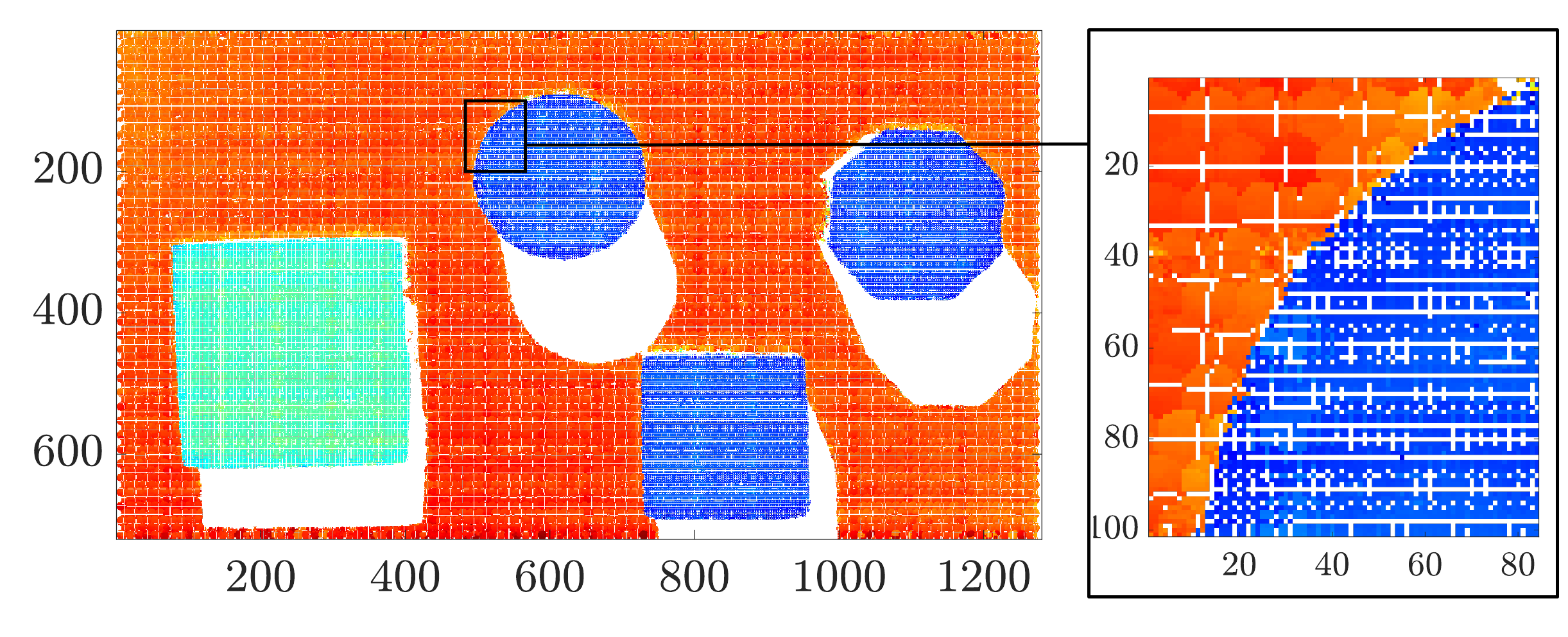}                
            }
            \\
            \subfloat[]
            {
                \centering
                \includegraphics[width=0.4\linewidth]{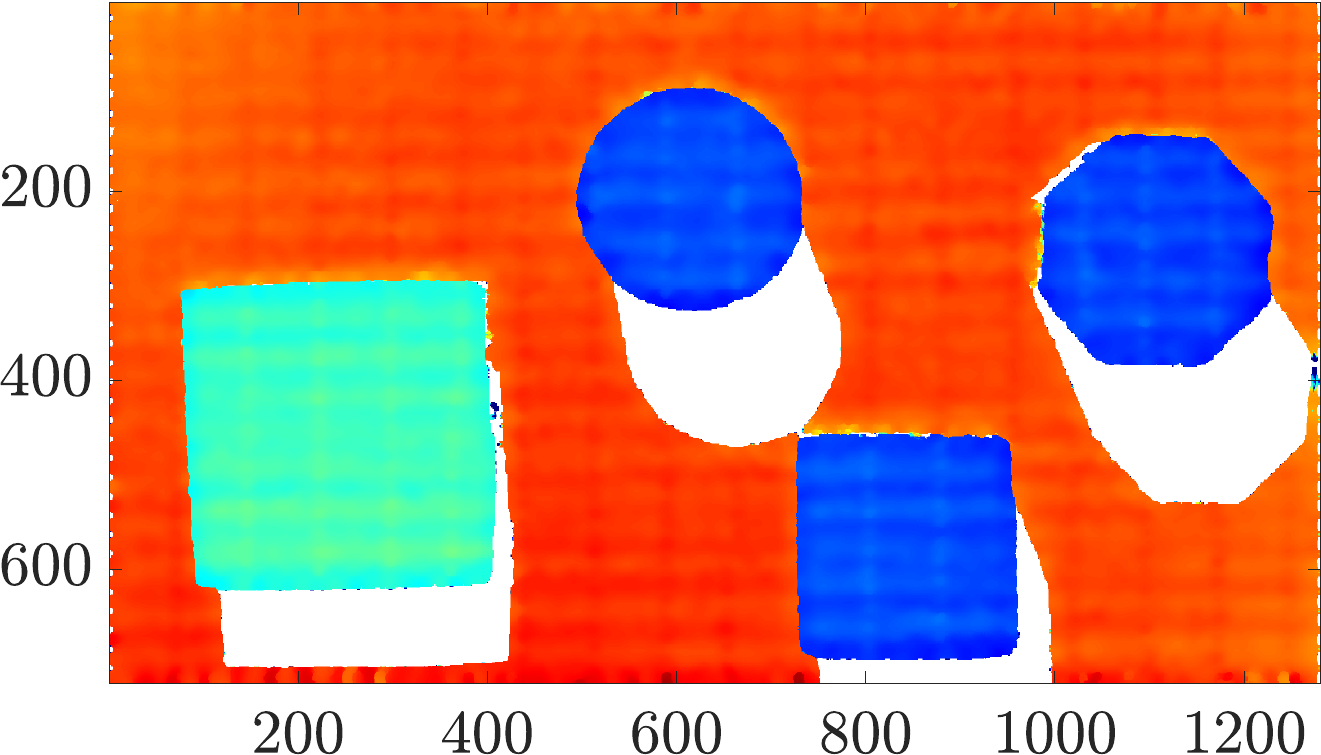}
            }		
            \subfloat[]
		{
			\centering
			\includegraphics[width=0.4\linewidth]{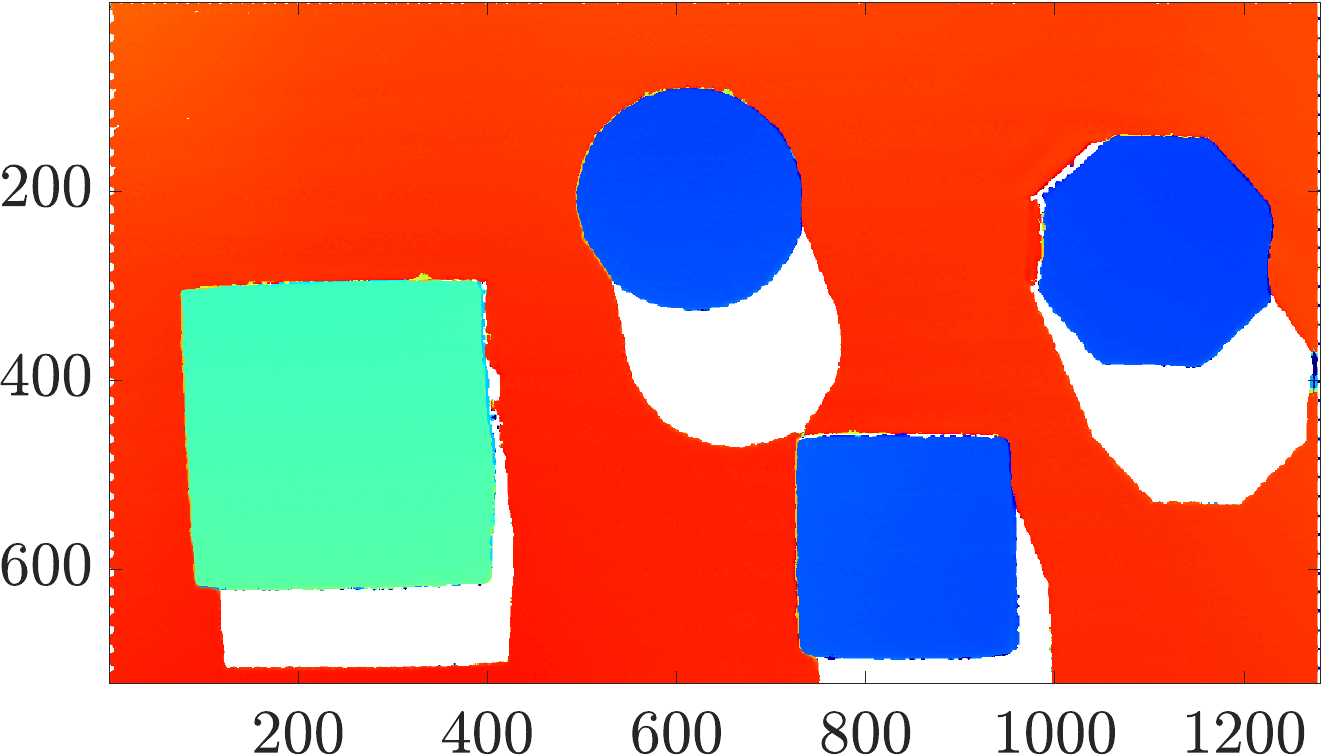}
		}
            \\
            \subfloat[]
		{
			\centering
			\includegraphics[width=0.15\linewidth]{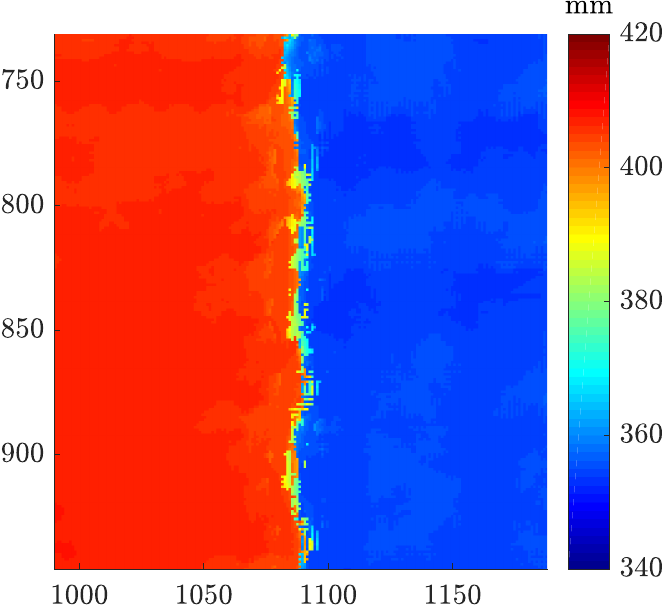}
		}
		\subfloat[]
		{
			\centering
			\includegraphics[width=0.15\linewidth]{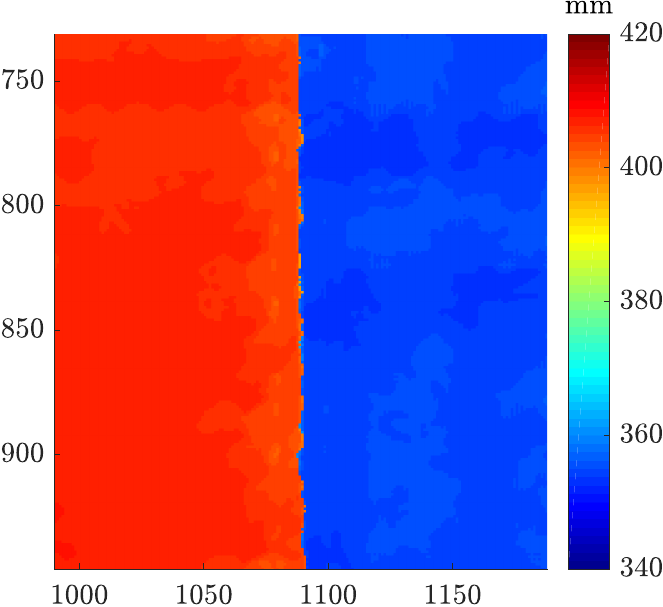}
		}
		\subfloat[]
		{
			\centering
			\includegraphics[width=0.15\linewidth]{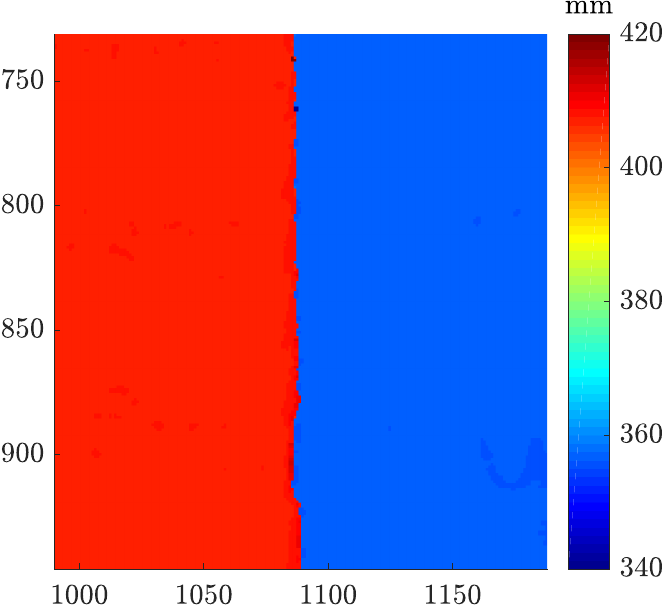}
		}
            \caption{ Re-projection and refinement strategy for accurate point clouds. (a) The initial depth map, (b) re-projected depth map, (c) reference depth map, and (d) final depth map. Our PSAD exhibits more robust performance at leapping area of the depth map: (e) classical SAD, (f) PSAD, and (g) ground truth.} 
		\label{FIG:FigDepth}
	\end{figure}
In order to reconstruct accurate point clouds, we propose a re-projection and refinement strategy in this section. First, we generate a reference depth map from the initial point clouds by re-projection, interpolation, and filtering, as shown in Fig.~\ref{FIG:FigDepth}(c); second, we refocus the captured fringe images according to the stereo matching results and compute the wrapped phase; finally, we compute the phase order to unwrap the phase according to the reference depth map, and compute the accurate 3D point clouds with the absolute phase, as shown in Fig.~\ref{FIG:FigDepth}(d). Implementation details in this section can be found in the supplementary material.

\section{Experimental Results}
	\subsection{Configuration}
	The proposed algorithm is implemented on an AMD Ryzen 9 7945HX @ 2.50 GHz with 16 GB RAM and written in MATLAB. Our experimental system consists of a Raytrix R8 plenoptic camera with angular resolution of $35\times35$ and spatial resolution of $110\times71$ (imaging on a $3840\times2160$ resolution sensor), and a TI DLP4500 projector with a resolution of $912\times1140$. The parameter settings of our algorithm are clarified in section 3.
 
 \subsection{Qualitative Evaluations}
 
     \begin{figure*}[t]
    		\centering
                \includegraphics[width=0.95\linewidth]{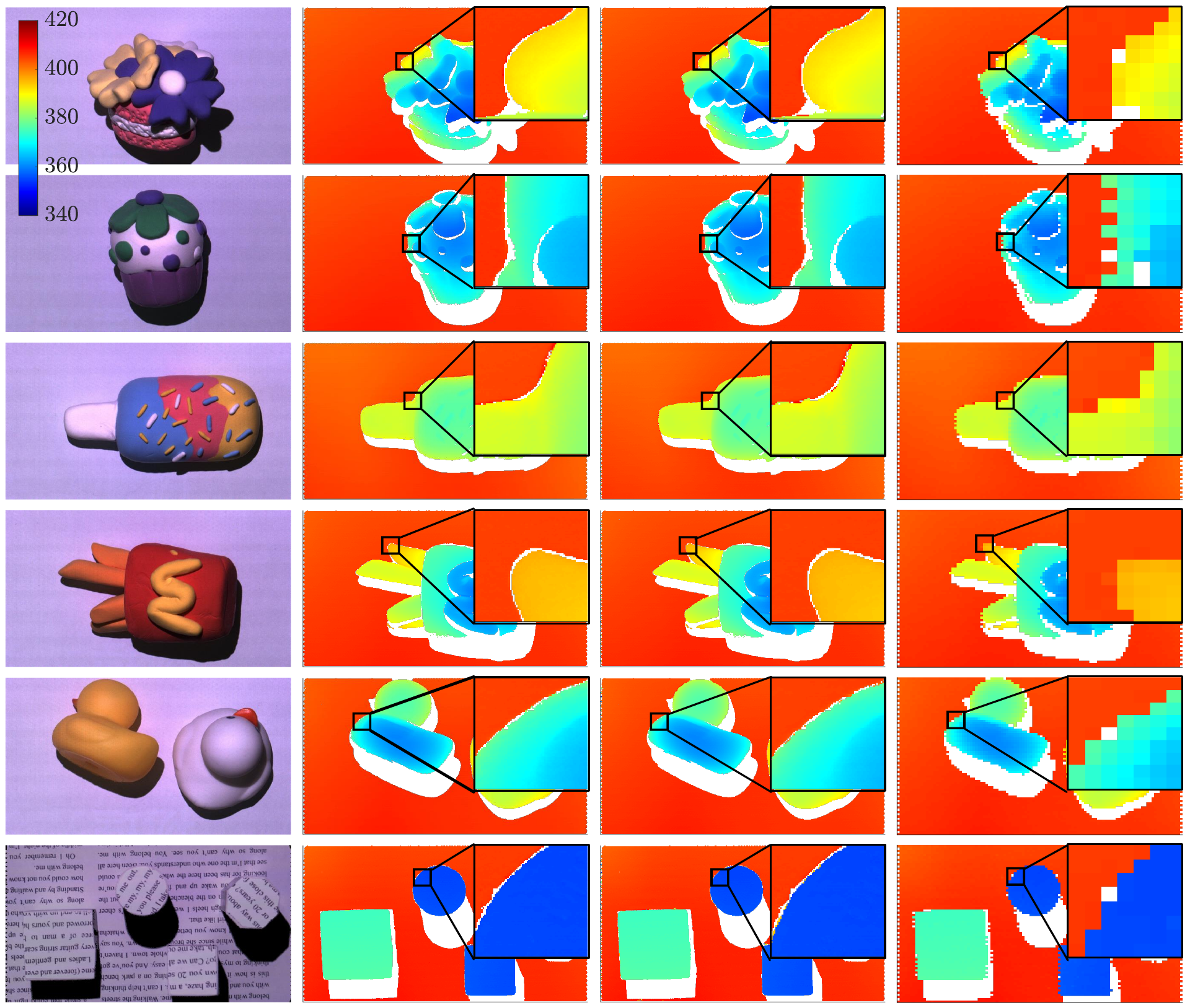}
    		\caption{Visualization of the reconstruction results. \textbf{Column 1}: measured object, \textbf{Column 2}: ground truth, \textbf{Column 3}: our final result~(resolution: 1280$\times$720), \textbf{Column 4}: state-of-the-art active light field techniques~\cite{cai2016structured,cai2018light,cai2020structured}~(resolution: 110$\times$71). Our method exhibits clear and sharp depth at discontinuities, while existing methods exhibit a mosaic-like pattern due to their lower spatial resolution. Therefore, our method achieves a stronger resolving ability than existing active light field techniques.}
    		\label{FIG:FigRecontructionResult}
    \end{figure*}
    
In the first group of experiments, we employed our method and state-of-the-art active light field techniques to reconstruct several clay handicrafts, thereby comparing the resolving ability of the depth map. The ground truth is obtained by traditional multi-frequency temporal phase shifting profilometry~\cite{yalla2005very}. The results are shown in Fig.~\ref{FIG:FigRecontructionResult}, our method exhibits clear and sharp depth at discontinuities, while existing methods exhibit a mosaic-like pattern due to their lower spatial resolution. Therefore, our method achieves a stronger resolving ability than existing active light field techniques.
 
 % The comparison of our PGLF, active light field, and passive light field in terms of accuracy, acquisition speed, and resolution of the depth map is qualitatively given in Tab.~\ref{TAB:TabQualitativeComparasion}.
 
 Our method only utilizes a single group of fringe patterns to have the equivalent accuracy as the multiple-shot active light field techniques, while achieving a significantly improved resolution of depth map. A visualized comparison with state-of-the-art active light field methods~\cite{cai2016structured,cai2018light,cai2020structured} is demonstrated in the supplementary materials.

\subsection{Quantitative Evaluations}	
 
	\begin{figure*}[t]
		\centering
            \includegraphics[width=0.95\linewidth]{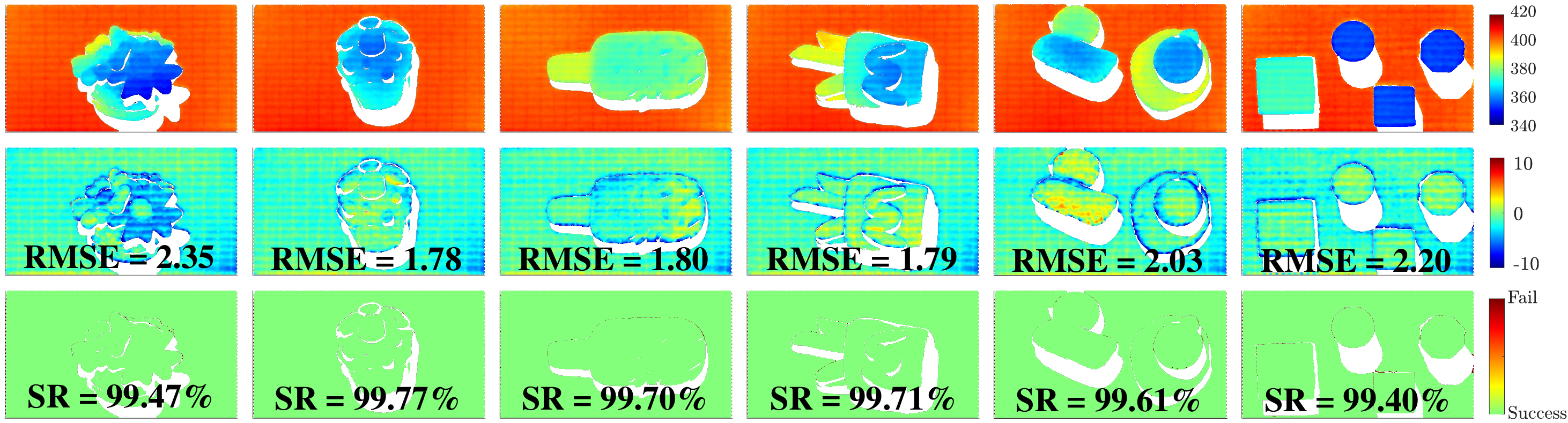}
		\caption{Visualization of the reconstruction results (resolution: 1280$\times$720). \textbf{Row 1}: reference depth obtained in section 5, \textbf{Row 2}: error of reference depth (mm), and \textbf{Row 3}: SR of unwrapping in final result.}
		\label{FIG:FigReferenceDepth}
	\end{figure*}
 
    In the first set of experiments, we quantitatively evaluate 1) reference depth: the performance of our DCM for suppressing the depth deformation caused by axial aberration; 2) final point clouds: the absolute accuracy of our final result. We employed our method with linear LF imaging model and the proposed DCM to reconstruct several clay handicrafts, with the frequency of $f = 32$ and phase shifting step number of $N = 6$, respectively, and simultaneously obtained the ground truth point cloud with multi-frequency temporal phase shifting profilometry, with the frequency of $f = \{1, 8, 32\}$ and phase shifting step number of $N = \{3, 3, 6\}$. The 3D reconstruction results are presented in Fig.~\ref{FIG:FigRecontructionResult} and were analyzed in qualitative evaluation. 
 	\begin{table}[h!]
            \small
		\renewcommand\arraystretch{1.1}
		\centering
		\caption{ Measurement results of handicrafts (units: mm).}
		\begin{tabular}{m{0.8cm}<{\centering}|m{1cm}<{\centering}m{1cm}<{\centering}m{1cm}<{\centering}|m{1cm}<{\centering}m{1cm}<{\centering}m{1cm}<{\centering}}
			\hline
			\multirow{3}[4]{*}{no.} & \multicolumn{3}{c|}{Linear LF model} & \multicolumn{3}{c}{Our DCM} \\
			\cline{2-7}          & \multicolumn{2}{c|}{Reference depth} & \multirow{2}[0]{*}{SR} & \multicolumn{2}{c|}{Reference depth} & \multirow{2}[0]{*}{SR} \\
			\cline{2-3}\cline{5-6}          & RMSE   & \multicolumn{1}{c|}{MAE} &       & RMSE   & \multicolumn{1}{c|}{MAE} &  \\
			\hline
		    1     & \multicolumn{1}{c}{8.158 } & \multicolumn{1}{c}{7.197 } & 81.47\%  & \multicolumn{1}{c}{1.991} & \multicolumn{1}{c}{1.583 } & 99.61\%  \\
			2     & \multicolumn{1}{c}{7.706 } & \multicolumn{1}{c}{6.517 } & 90.12\%  & \multicolumn{1}{c}{ 1.409} & \multicolumn{1}{c}{1.094} & 99.72\%  \\
			\hline
			Mean  & \multicolumn{1}{c}{ 7.932}  & \multicolumn{1}{c}{6.857}  & 85.80\%  & 1.700  &\multicolumn{1}{c}{1.339} & 99.67\%  \\
			\hline
		\end{tabular}%
		\label{TAB:TabErrorHandicraft}
	\end{table}% 

    The accuracy of the reference depth map determines the success rate (SR) of phase unwrapping. Therefore, we first to quantitatively assess the accuracy of the reference depth map, we employ the root mean squared error (RMSE) and mean absolute error (MAE) as metrics for evaluating reference depth. We also use the success rate (SR) of unwrapping for the final point clouds as an evaluation criterion. The results are shown in Fig.~\ref{FIG:FigReferenceDepth}. It can be seen that the reference depth is of reliable accuracy to ensure a high SR of phase unwrapping, thereby reducing the outliers in the final point clouds.
	
    Then we slightly adjust the focal distance of the plenoptic camera and measure the clay handicrafts again, the results of the two groups of measurements are listed in Tab.~\ref{TAB:TabErrorHandicraft}. It is evident that when compared with the linear LF imaging model, the error in reference depth obtained by our DCM is significantly reduced in RMSE and MAE, consequently a large area of unwrapping failed region is eliminated, so the SR of unwrapping is improved from 85.80$\%$ to 99.67$\%$. We conclude that our DCM can effectively suppress the axial aberration of the main lens of a plenoptic camera. Moreover, our PSAD shows robust performance over depth-discontinuous areas.

    \begin{figure}[!h]
    \centering
      \begin{minipage}{0.28\linewidth}
            \centering
            \includegraphics[width=\linewidth]{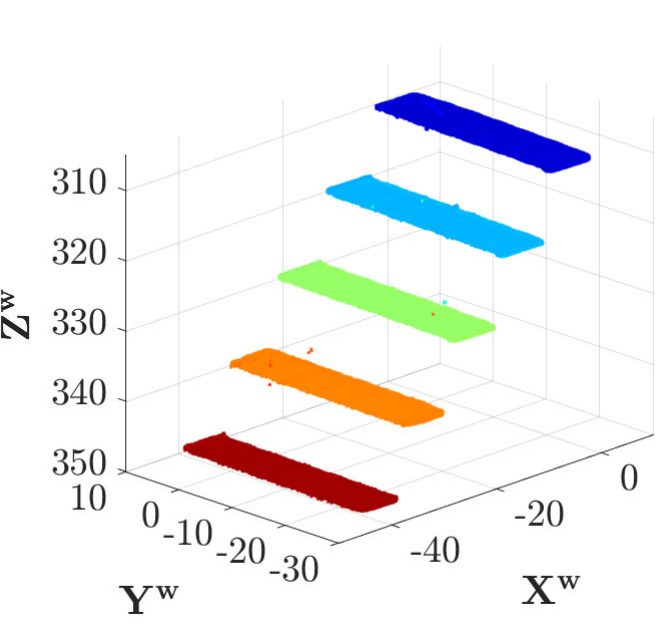}
      \end{minipage}
     \begin{minipage}{0.5\linewidth}
            \footnotesize
                \centering
                \setlength{\tabcolsep}{8pt} 
    		\begin{tabular}{c|cccc}
    			\hline
    			no. & GT & Mean  & Error & STD \\
    			\hline
    			1 & 10    & 9.9677  & 0.0323  & 0.0787  \\
    			2 & 10    & 9.9504  & 0.0496  & 0.0590  \\
    			3 & 10    & 9.9219  & 0.0781  & 0.0585  \\
    			4 & 10    & 9.8384  & 0.1616  & 0.0501  \\
    			\hline
    			Mean  &  10   & 9.9196  & 0.0804  & 0.0616  \\
    			\hline
    	\end{tabular}
      \end{minipage}
        \caption{The measured result of gauge blocks (units: mm).} 
        \label{FIG:FigGaugeBlock}
    \end{figure}  
    
    The second set of experiments aims at assessing the absolute error of our 3D reconstruction method. We conduct the following measurements at a distance of 400~mm: 1) step height measurement: we measure the step height of five gauge blocks arranged in tiers, each with a known ground truth value of 1cm. The results of the measured step height of the staircase are visualized in Fig.~\ref{FIG:FigGaugeBlock}. 2) circle center distance measurement: we measure a calibration board featuring a grid of $6\times7$ circular identifiers, and compute the circle center distance according to the measured point cloud. The fringe patterns are of settings $f = 32$ and $N = 8$. The results are listed in Tab.~\ref{TAB:TabCircleCenterDistance}.	
    \setlength{\tabcolsep}{8pt} 
	\begin{table}[!h]
            \small
		\renewcommand\arraystretch{1}
		\centering
		\caption{Measurement results of circle center distance (units: mm).}
		\begin{tabular}{ccccc}
			\hline
			GT & Mean  & MAE & STD & Percentage MAE \\
			\hline
			50    & 49.9492  & 0.0508 & 0.0436  & 0.1016\%  \\
			\hline
		\end{tabular}%
		\label{TAB:TabCircleCenterDistance}
	\end{table}%
    The experimental results show that, our PGLF accomplishes dual-high measurements on spatial and depth with resolution of $1280\times720$ and standard derivation of less than 70~$\mu$m. Therefore, we conclude that our method is capable of industrial-grade 3D measurements. It is notable that the MAEs reach 0.0804~mm and 0.0508~mm for measuring step height and circle center distance, respectively. The MAE can primarily be attributed to distortions caused by the main lenses in the plenoptic camera and projector, which can be addressed by well-documented lens distortion correction methods~\cite{zhang2000flexible,moreno2012simple,zhang2022correcting}. 

% Moreover, we believe that the standard deviation is induced by 1) bayer array: the bayer array sensors decrease the ERR of the sensors, 2) refocus algorithm: our refocusing algorithm directly averages all the intensities of the corresponding pixel without selecting the most suitable type of lenslets, 3) fringe frequency: we intentionally did not set the fringe frequency too high, because the field of view of our projector is excessively wider than the plenoptic camera, and too high a frequency induces a significant decrease of the modulation $B^c$ and reduce the signal-to-noise ratio. These challenges can be tackled by improving either the hardware or algorithm, which means our PGLF can theoretically achieve a higher accuracy.

    The proposed algorithm required about 80s for each reconstruction procedure. It is worth noting that this algorithm is implemented using CPU multi-threading (10 threads) in MATLAB without any optimization. Therefore, it is anticipated that a substantial improvement in speed could be achieved by parallelizing stereo matching through GPU utilization.    

\section{Conclusions and Future Work}
    In summary, we proposed a PGLF pipeline for reconstructing spatial-depth high resolution and unambiguous 3D point clouds from a merely single group of high-frequency fringe patterns, implemented in mainstream plenoptic camera 2.0 architecture. We conclude that our method's dual-high resolution enables LF 3D imaging to achieve dense and precise industrial-grade 3D measurements, and the reduction in the number of fringe patterns significantly improves scanning time efficiency. 
	
	However, there are several existing issues that warrant future consideration. Although our PSAD effectively overcomes depth-discontinuous issue in most cases, the result is unsatisfactory when the depth difference exactly induce a phase difference of integer multiple periods. Another issue of the proposed is the reliance on graphic process units in practical applications due to the high time consumption.

\bibliographystyle{splncs04}
\bibliography{main}
\end{document}

\title{Supplementary Material of "Phase-Guided Light Field for Spatial-Depth High Resolution 3D Imaging" } 

\titlerunning{Abbreviated paper title}

\author{First Author\inst{1}\orcidlink{0000-1111-2222-3333} \and
Second Author\inst{2,3}\orcidlink{1111-2222-3333-4444} \and
Third Author\inst{3}\orcidlink{2222--3333-4444-5555}}

\authorrunning{F.~Author et al.}

\institute{Princeton University, Princeton NJ 08544, USA \and
Springer Heidelberg, Tiergartenstr.~17, 69121 Heidelberg, Germany
\email{lncs@springer.com}\\
\url{http://www.springer.com/gp/computer-science/lncs} \and
ABC Institute, Rupert-Karls-University Heidelberg, Heidelberg, Germany\\
\email{\{abc,lncs\}@uni-heidelberg.de}}

\maketitle
\begin{appendices}
\section{Imaging principle of plenoptic camera}
	\begin{figure}[!h]
		\centering
		\includegraphics[width=0.7\linewidth]{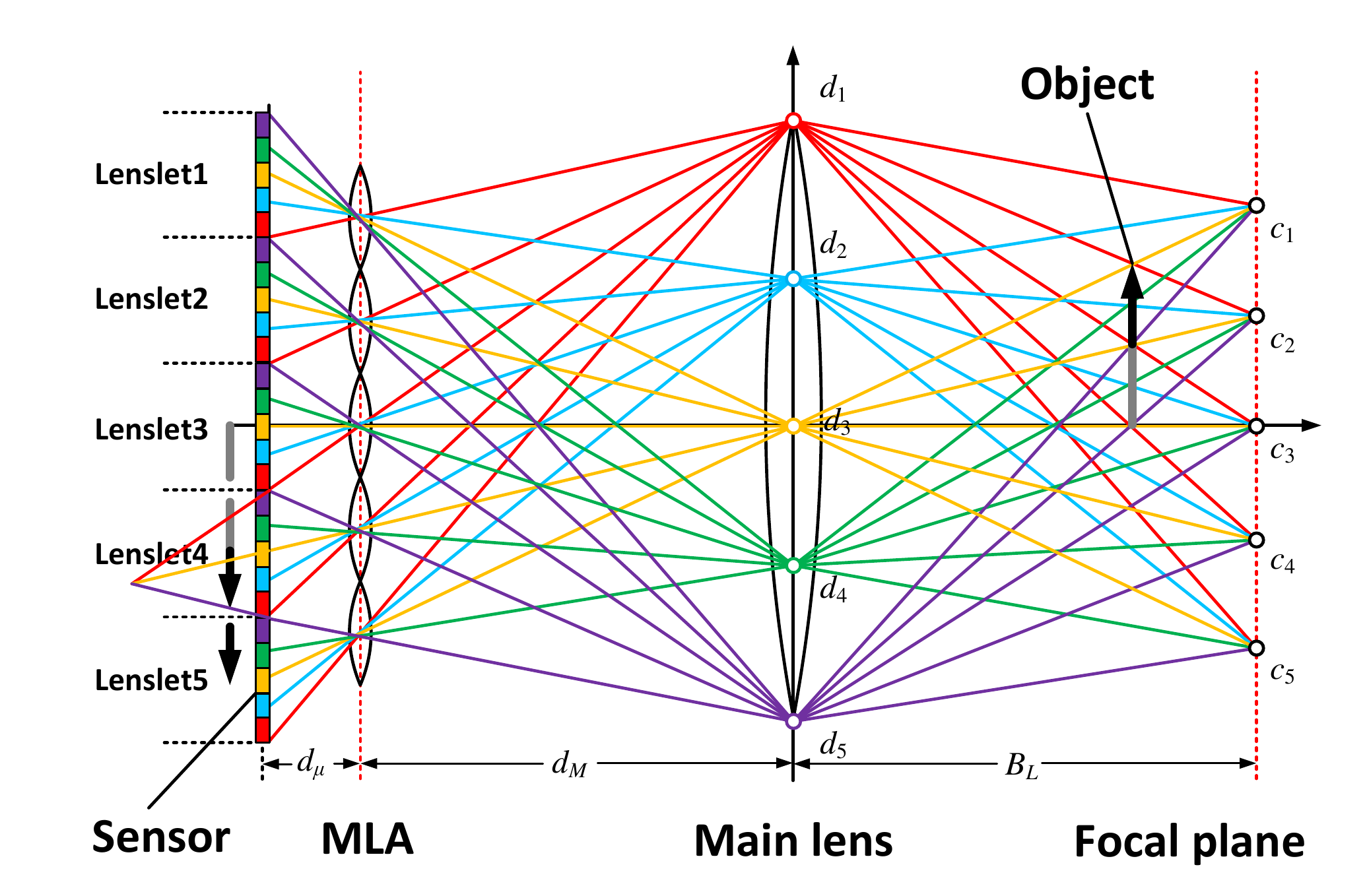}
		\caption{Imaging principle of a focused plenoptic camera.}
		\label{Fig:FigImagingPrinciple}
	\end{figure}
    The imaging principle of a focused plenoptic camera is shown in Fig.~\ref{Fig:FigImagingPrinciple}. From one perspective, $\{c_{i}\}$ (where $i={1,2,3,4,5}$) on the focal plane can be considered as the optical centers of a lenslet camera array, and each lenslet camera has the same resolution as the lenslet image; from another perspective, $\{d_{i}\}$ (where $i={1,2,3,4,5}$) on the main lens plane can be viewed as the optical centers of a sub-aperture camera array. 
    
\section{Fringe Frequency Selection}
	By projecting high-frequency fringe patterns onto the object, we actively cover the target with periodic textures, i.e., wrapped phase, which can be regarded as repetitive but periodically controllable texture. In this paper, we conduct a stereo matching between the phase map of lenslets and generate a reference depth map for high-frequency phase unwrapping, which can significantly reduce the number of patterns and improve the accuracy of light field 3D imaging to a structured-light level. To ensure the correctness of stereo matching and phase unwrapping, we propose an uniqueness constraint to prevent ambiguity in stereo matching caused by repeating phase values.
	
	\subsubsection{Phase Sensitivity}
	The corresponding phase of a 3D point $(X^w, Y^w, Z^w)$ can be computed as
	\begin{equation}\label{EQ:3DPoint2Phase}
		{y^p} = \frac{{m_{21}^p{X^w} + m_{22}^p{Y^w} + m_{23}^p{Z^w} + m_{24}^p}}{{m_{31}^p{X^w} + m_{32}^p{Y^w} + m_{33}^p{Z^w} + m_{34}^p}}.
	\end{equation}
	The quantitative calculation of the two constraints depends on the mathematical relationship between the phase difference and 3D difference. Thus, by computing the partial derivative of $y^p$ with respect to $X^w$, $Y^w$, and $Z^w$, we derive the phase sensitivity as
	\begin{equation}\label{EQ:3DSensitivity}
		\left\{
		\begin{split}
			\frac{\partial y^p }{\partial {X^w}} = \frac{{{q_1}{X^w} + {r_1}{Y^w} + {s_1}{Z^w} + {t_1}}}{{{{\left( {m_{31}^p{X^w} + m_{32}^p{Y^w} + m_{33}^p{Z^w} + m_{34}^p} \right)}^2}}}\\
			\frac{\partial y^p }{\partial {Y^w}} = \frac{{{q_2}{X^w} + {r_2}{Y^w} + {s_2}{Z^w} + {t_2}}}{{{{\left( {m_{31}^p{X^w} + m_{32}^p{Y^w} + m_{33}^p{Z^w} + m_{34}^p} \right)}^2}}}\\
			\frac{\partial y^p }{\partial {Z^w}} = \frac{{{q_3}{X^w} + {r_3}{Y^w} + {s_3}{Z^w} + {t_3}}}{{{{\left( {m_{31}^p{X^w} + m_{32}^p{Y^w} + m_{33}^p{Z^w} + m_{34}^p} \right)}^2}}}
		\end{split}\right.,
	\end{equation}
	where
	\begin{equation}\nonumber
		\begin{gathered}
			\left\{
			\begin{split}
				q_1 = {m_{21}^pm_{31}^p - m_{31}^pm_{21}^p}\\
				r_1 = {m_{21}^pm_{32}^p - m_{31}^pm_{22}^p}\\
				s_1 = {m_{21}^pm_{33}^p - m_{31}^pm_{23}^p}\\
				t_1 = {m_{21}^pm_{34}^p - m_{31}^pm_{24}^p}\\
			\end{split}\right.,
			\left\{
			\begin{split}
				q_2 = {m_{22}^pm_{31}^p - m_{32}^pm_{21}^p}\\
				r_2 = {m_{22}^pm_{32}^p - m_{32}^pm_{22}^p}\\
				s_2 = {m_{22}^pm_{33}^p - m_{32}^pm_{23}^p}\\
				t_2 = {m_{22}^pm_{34}^p - m_{32}^pm_{24}^p}\\
			\end{split}\right., \\
			\mathrm{and}
			\left\{
			\begin{split}
				q_3 = {m_{23}^pm_{31}^p - m_{33}^pm_{21}^p}\\
				r_3 = {m_{23}^pm_{32}^p - m_{33}^pm_{22}^p}\\
				s_3 = {m_{23}^pm_{33}^p - m_{33}^pm_{23}^p}\\
				t_3 = {m_{23}^pm_{34}^p - m_{33}^pm_{24}^p}\\
			\end{split}\right..
		\end{gathered}
	\end{equation}
	Once the calibration of the camera-projector system is completed~\cite{yalla2005very}, we can conveniently compute the phase sensitivity at arbitrary 3D coordinates, because $\frac{\partial y^p }{\partial {X^w}}$, $\frac{\partial y^p }{\partial {Y^w}}$, and $\frac{\partial y^p }{\partial {Z^w}}$ are functions of $(X^w, Y^w, Z^w)$.
	\subsubsection{Uniqueness Constraint}
	\begin{figure}[!h]
		\centering
		\includegraphics[width=1\linewidth]{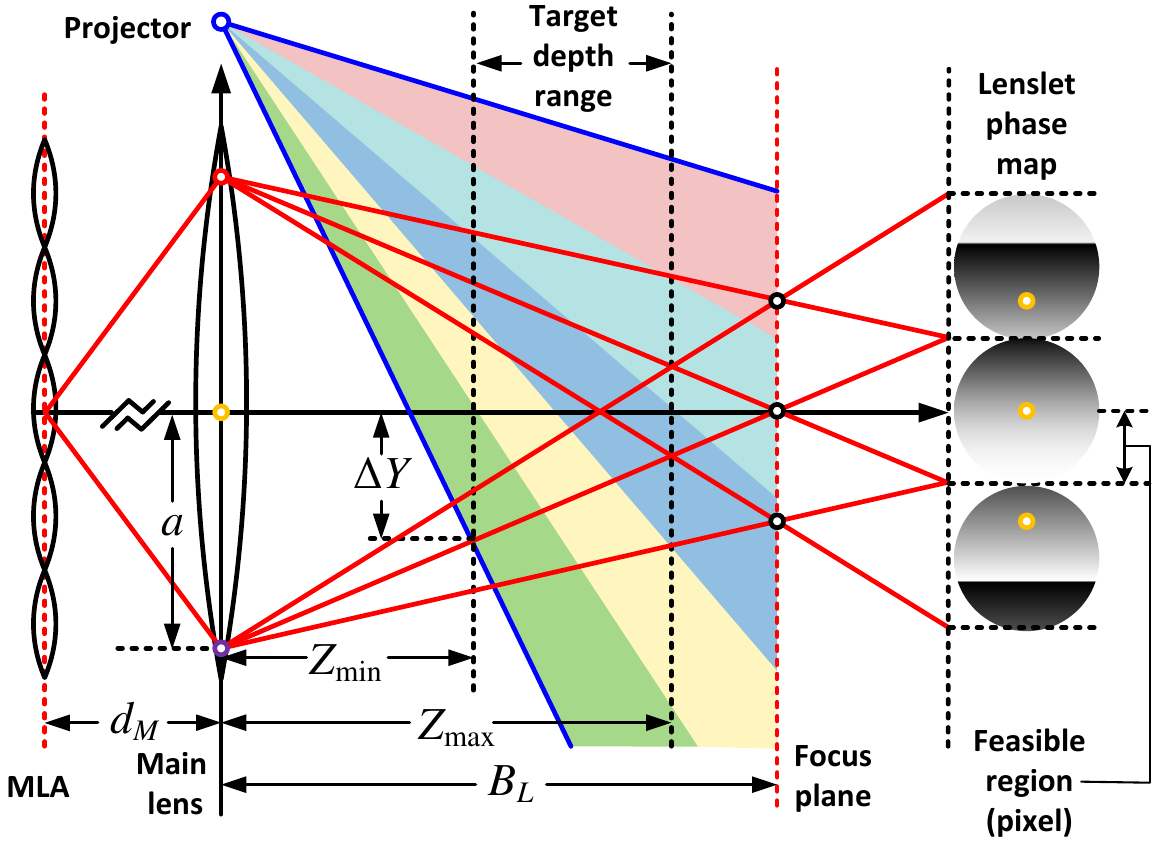}
		\caption{Imaging principle of a focused plenoptic camera.}
		\label{Fig:FigUniquenessConstraint}
	\end{figure}
	To prevent ambiguity in stereo matching, the phase value in the feasible region of matching should be unique as shown in Fig.~\ref{Fig:FigUniquenessConstraint}, i.e., the feasible region can only contain one period of phase at most. Technically, the feasible region in lenslet images is determined by the target depth range which can be obtained from actual conditions. However, in order to leave enough margin, we assume the feasible region as the radius of the lenslet image as shown in Fig.~\ref{Fig:FigUniquenessConstraint}.
	
	For the purpose of obtaining the fringe frequency constraint, first, we compute the feasible region width at lower limit of target depth range as
	\begin{equation}
		\Delta Y = \left(1 - \frac{Z^w_\mathrm{min}}{B_L}\right)a,
	\end{equation}
	where $Z^w_\mathrm{min}$ is the lower limit of target depth range, $B_L$ is the distance between the main lens plane and the focus plane of MLA, and $a$ is half the aperture diameter; Second, based on 3D sensitivity, we can compute the phase difference in the projector space corresponding to the feasible region width
	\begin{equation}\label{EQ:DeltaY}
		\Delta y^p = \left|\frac{\partial y^p }{\partial {Y^w}} \Delta Y \right|,	
	\end{equation}
	where ${\left( {{X^w},{Y^w},{Z^w}} \right) = \left( {\frac{{Z^w_\mathrm{min}}}{{{f_x}}}{x^c},\frac{{Z^w_\mathrm{min}}}{{{f_y}}}{y^c},Z^w_\mathrm{min}} \right)}$ in $\frac{\partial y^p }{\partial {Y^w}}$, $(x^c,y^c)$ is the integer-valued camera space coordinate, $f_x$ and $f_y$ are the focal length of the camera in pixel. Note that we omit the very small $\frac{\partial y^p }{\partial {X^w}}\Delta X$ and $\frac{\partial y^p }{\partial {Z^w}}\Delta Z$ terms in Eq.~(\ref{EQ:DeltaY}).
	
%	Fig.~\ref{} gives the phase difference map, which shows the phase change corresponding to $\Delta Y$ change of $Y$ coordinate at the 3D point $(\frac{Z^w_\mathrm{min}}{f_x}x^c, \frac{Z^w_\mathrm{min}}{f_y}y^c, Z^w_\mathrm{min})$. 
	
	We numerically find the maximum value of the phase difference map
	\begin{equation}
		\Delta y^p_{\mathrm{max}} = \mathop {\max }\limits_{\left\{\left(x^c,y^c\right)\right\}}{\left(\Delta y^p\right)},	
	\end{equation}
	and take $\Delta y^p_{\mathrm{max}}$ as the longest period to compute the upper limit of frequency of fringes as 
	\begin{equation}\label{EQ:UniquenessConstraint}
		f < \frac{H^p}{\Delta y^p_{\mathrm{max}}}.	
	\end{equation}
	For more effectively suppressing the noises on phase, we set the fringe frequency as high as possible on the premise that Eq.~(\ref{EQ:UniquenessConstraint}) is satisfied.

\section{Feasible Region Determining}
    \begin{figure}[!h]
            \centering
            \includegraphics[width=0.4\linewidth]{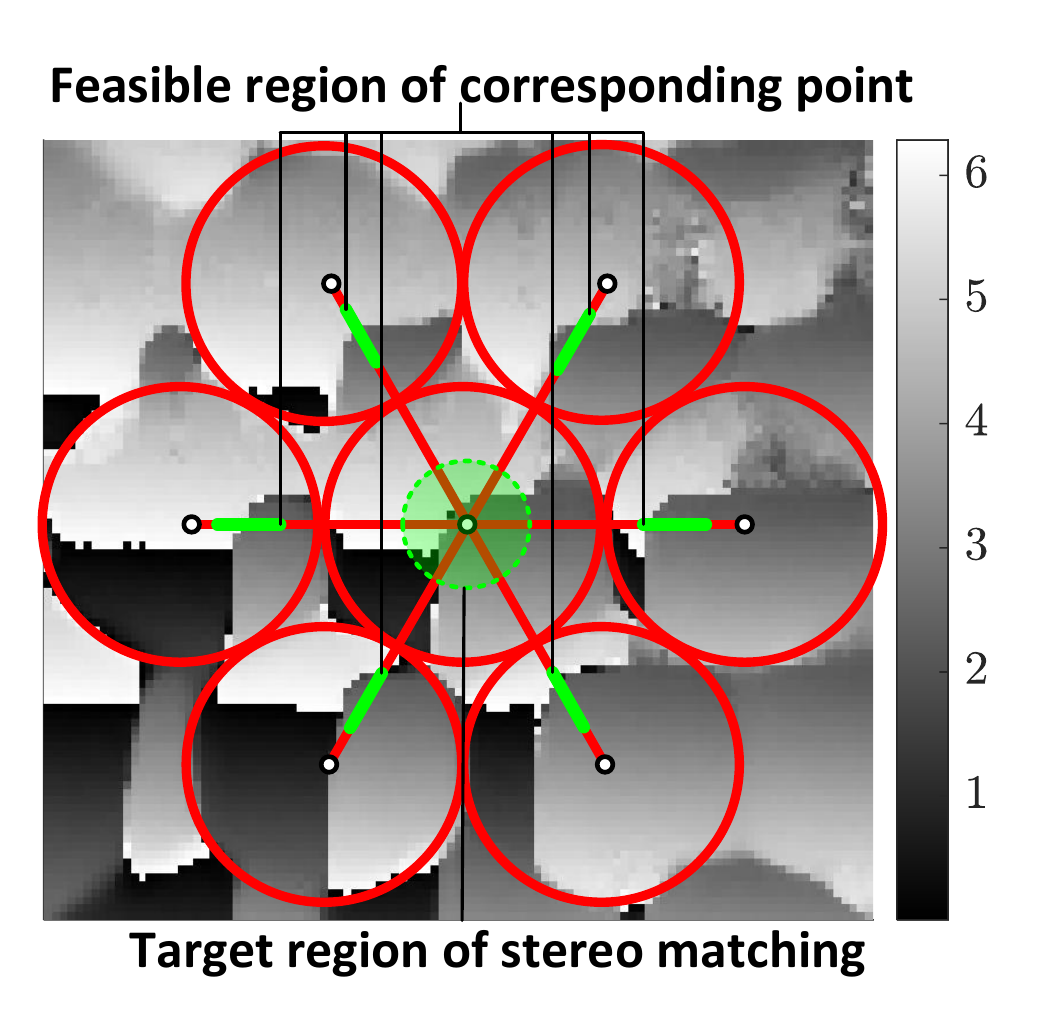}
        \caption{Epipolar constraint of plenoptic camera in a 2D view.} 
        \label{FIG:FigEpipolarConstraint}
    \end{figure}
    Before we conduct the phase based stereo matching, the feasible region of corresponding point distance should be determined to reduce the computational complexity. We analytically relate corresponding point distance $D$ to depth $Z^w$ as
     \begin{equation}\label{EQ:FeasibleRegion}
		\left\{ \begin{split}
			D &= \left( {1 - \frac{1}{v}} \right){D_\mu }\\
			v &= \frac{{ - \left(d_\mu+kd\right) + \sqrt {\left(d_\mu+kd\right)^2 - 4k{d_\mu }c} }}{{2k{d_\mu }}}\\
			c &= d - \frac{{{Z^w}{f_L}}}{\left( {{Z^w} - {f_L}} \right){\textbf{a}^{\text{T}}}{\textbf{p}}}\\
		\end{split}  \right..
    \end{equation}
    Consequetly, once we estimate a rough prior depth range $\left[ Z^w_\mathrm{min},Z^w_\mathrm{max} \right]$, the feasible region~(the green line segments) of corresponding point distance $\left[ D_\mathrm{min},D_\mathrm{max} \right]$ can be determined as shown in Fig.~\ref{FIG:FigEpipolarConstraint}.

    Although we can compute the correspondence of any template point we want, it is not necessary to match all the points in each lenslet image due to the overlap of the lenslet's field of view. In practice, for roughly ensuring there is no overlap, we only compute the correspondence of pixels within a circularly-shaped non-overlapping imaging area around the lenslet center, and the radius of the circle can be estimated as
	\begin{equation}\label{EQ:SpeckleRadius}
		r = \frac{{d_\mu }{B_L}Z^w_\mathrm{max}{D_\mu}}{2{d}\left( {d + d_\mu } \right)\left( {B_L - Z^w_\mathrm{max}} \right)},
	\end{equation}
    whereby we determine the target region for implementing stereo matching as shown in Fig.~\ref{FIG:FigEpipolarConstraint}.
\section{Re-projection and Refinement Strategy for Accurate Point Clouds}	
\subsection{Initial 3D Point Clouds Computing}	
	\begin{figure}[!h]
		\centering
		\includegraphics[width=0.9\linewidth]{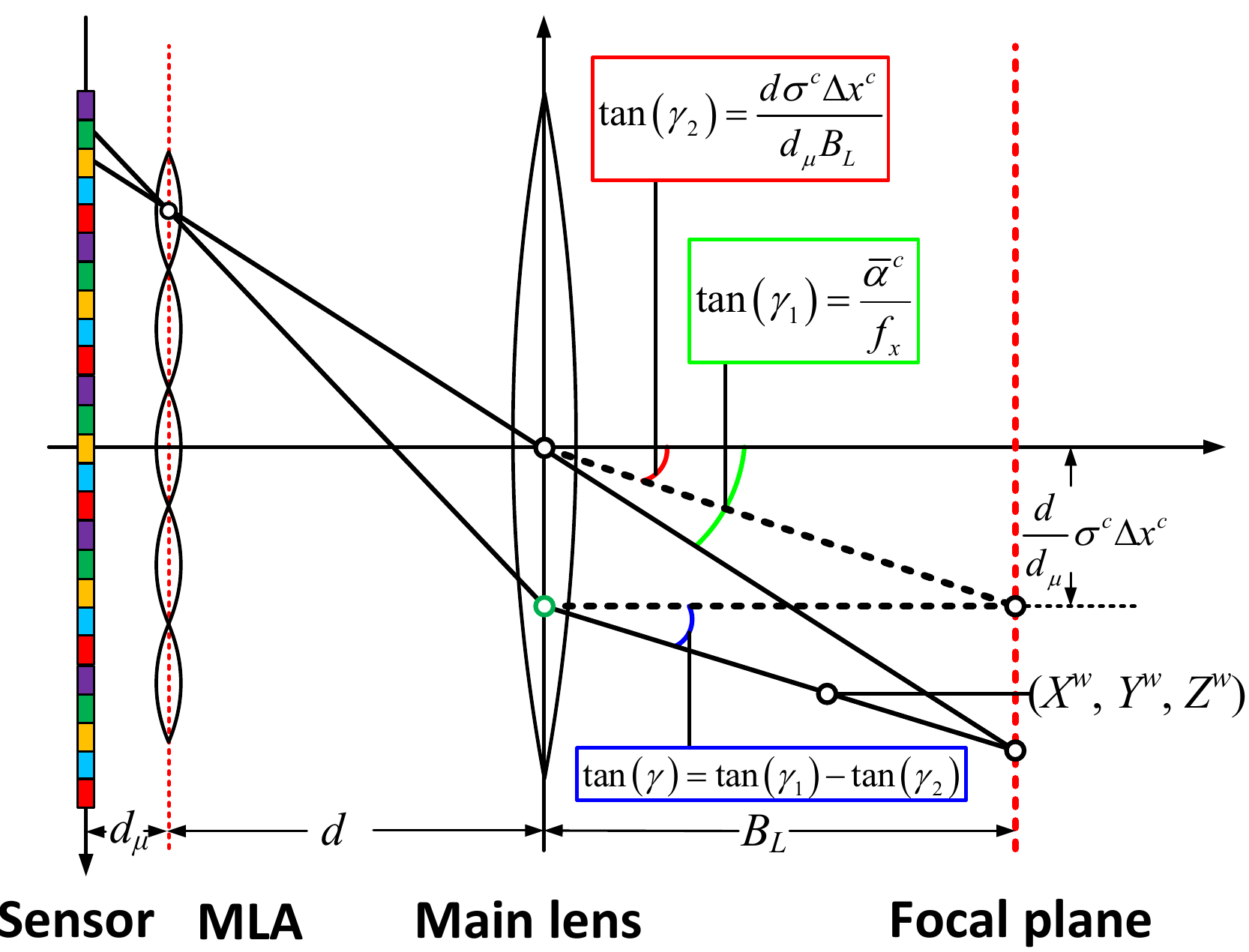}
		\caption{The geometrical optics relationship in a plenoptic camera.} 
		\label{FIG:FigXwYw}
	\end{figure}	
	For correcting the depth deformation with incident angle $(\theta_x, \theta_y)$, $X^w$ and $Y^w$ are needed. According to the geometrical optics in a plenoptic camera, we derive the expression as
	\begin{equation}\label{EQ:XwAndYw}
		\left[ \begin{gathered}
			{X^w} \hfill \\
			{Y^w} \hfill \\ 
		\end{gathered}  \right] = \emph{\textbf{k}}{Z^w} + \emph{\textbf{b}}
	\end{equation}
	where
	\begin{equation}\label{EQ:XwYw}
		\emph{\textbf{k}} = \left[ \begin{gathered}
			{\frac{{\bar \alpha^c}}{{{f_x}}} - \frac{{d{\sigma ^c}\Delta {x^c}}}{{{d_\mu }{B_L}}}} \hfill \\
			{\frac{{\bar \beta^c}}{{{f_y}}} - \frac{{d{\sigma ^c}\Delta {y^c}}}{{{d_\mu }{B_L}}}} \hfill \\ 
		\end{gathered}  \right]
		,\emph{\textbf{b}} = \frac{d\sigma^c}{{{d_\mu }}}\left[ \begin{gathered}
			\Delta {x^c} \hfill \\
			\Delta {y^c} \hfill \\ 
		\end{gathered}  \right],
	\end{equation}
    where $(\bar \alpha^c,\bar \beta^c)$ is the centralized piexl coordinate of lenslet center corresponding to the pixel to be computed, $\sigma^c$ is pixel size of the plenoptic camera, $\Delta {x^c} = x^c - \alpha^c$, and $\Delta {y^c} = y^c -\beta^c$. Derivation details can be found in the supplementary material.
	
    Then we compute the incident angle $(\theta_x, \theta_y)$ by Eq.~(5) in the paper, and subsequently substitute $(\theta_x, \theta_y, v)$ into Eq.~(4) in the paper to have corrected depth $\tilde{Z}^w$. Finally, we conveniently compute $\tilde{X}^w$ and $\tilde{Y}^w$ by reusing Eq.~(\ref{EQ:XwAndYw}) to have initial 3D point clouds.

\subsection{Initial point clouds reprojection}
	The reference depth map for phase unwrapping needs to be dense and aligned. However, the initial depth map is sparse and holds the misaligned pixel coordinates of the plenoptic image, so a reorganization strategy is conducted for addressing the issues.
	
	\begin{figure}[!h]
		\centering
		\includegraphics[width=0.99\linewidth]{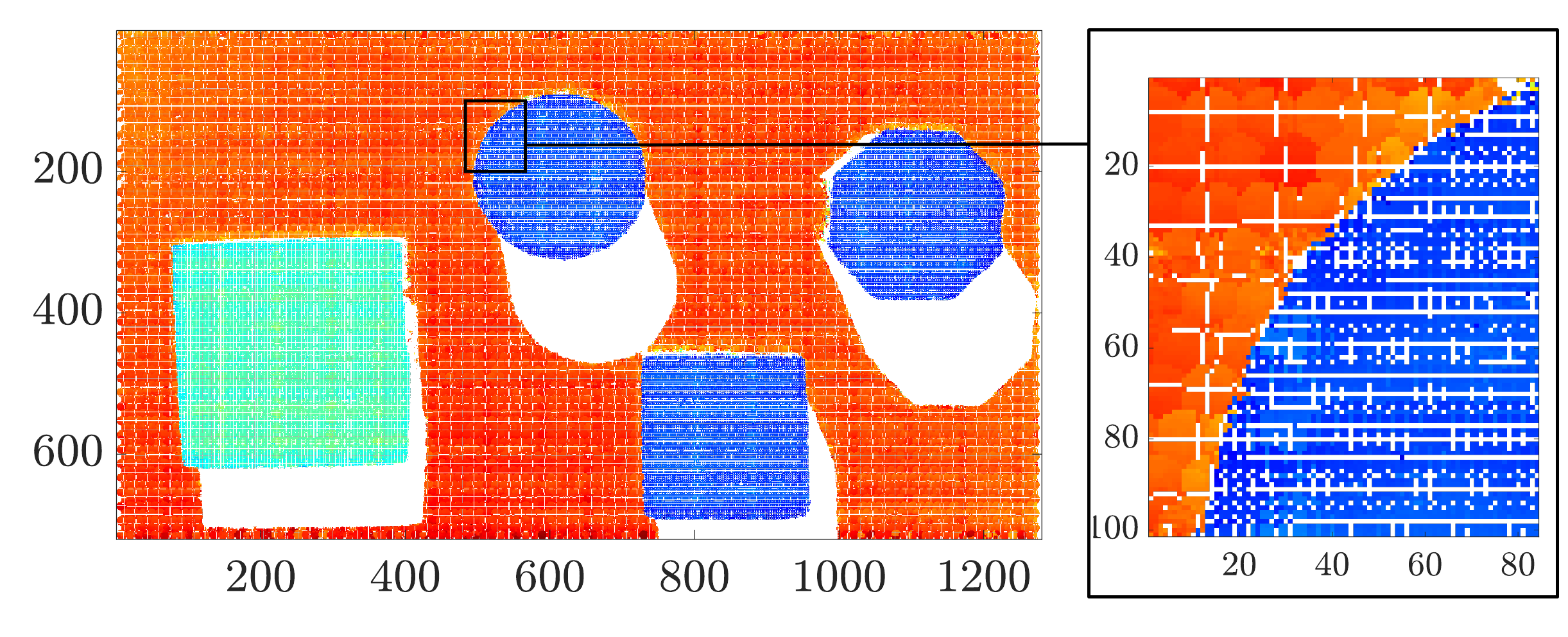}
		\caption{The reprojected depth map.} 
		\label{FIG:FigReprojectedDepthMap}
	\end{figure}
	
	We reproject the initial 3D points into the projector space to have an aligned but sparse depth map, as shown in Fig.~\ref{FIG:FigReprojectedDepthMap}. The smaller $Z^w$ is, the larger the speckles in the initial depth map will "diffuse" in the reprojected depth map. For ensuring that the points to be computed are as less as possible, by manipulating the radius of the speckles in Fig.~6(a) in the paper, the diffused circles will excatly touch each other at the depth of $D_\mathrm{max}$.
	
\subsection{Edge-preserved Interpolation of Depth Map}
\begin{figure}[!h]
		\subfloat[]
		{
			\centering
			\includegraphics[width=0.47\linewidth]{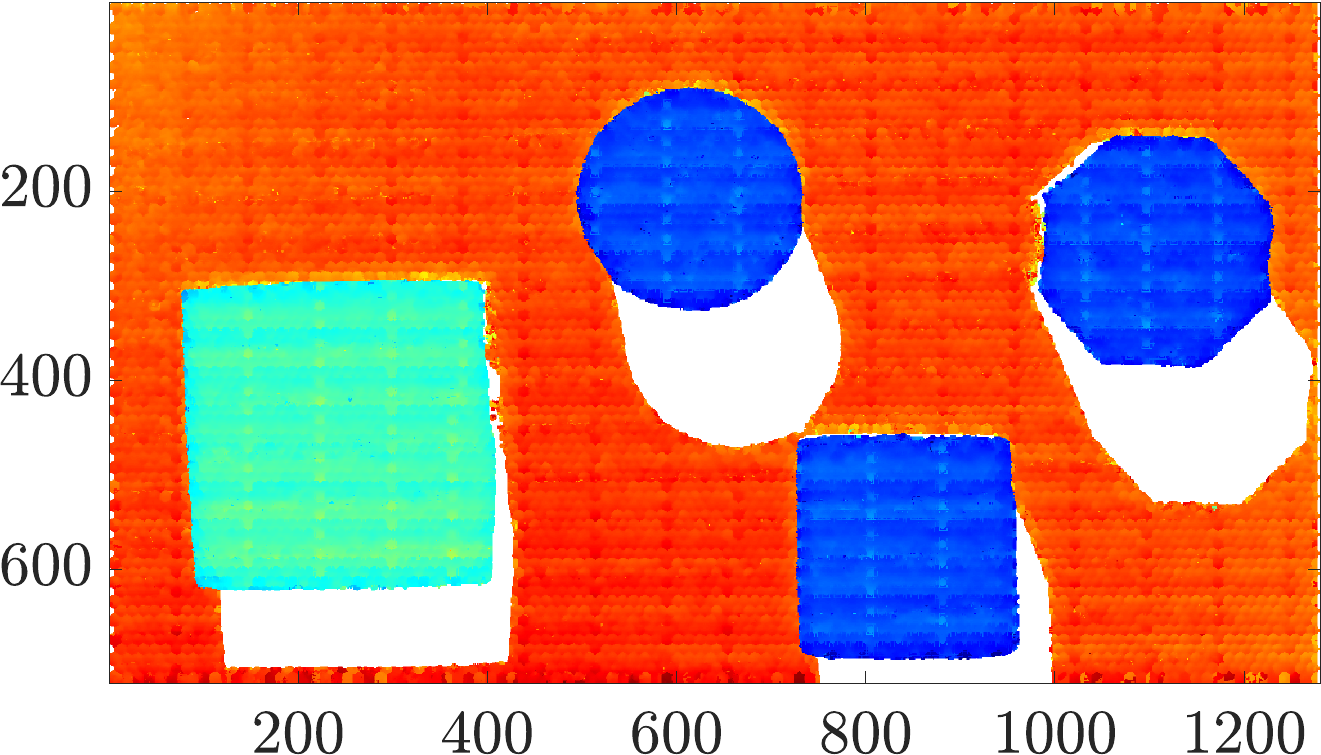}
		}
		\subfloat[]
		{
			\centering
			\includegraphics[width=0.47\linewidth]{./Figures/FigReferenceDepthMap}
		}
		\caption{The depth map (a) before and (b) after side window box filtering.} 
		\label{FIG:FigDenseDepthMap}
	\end{figure}

	We employ an edge-preserved interpolation to fill the blank space in the reprojected depth map.
	\begin{equation}\label{EQ:EdgePreservedInterpolation}
		{Z_r^w}\left( \textbf{s}^c \right) = \frac{1}{{\sum\limits_{\textbf{s}^c \in S} {U\left( \textbf{s}^c \right)} }}\sum\limits_{\textbf{s}^c \in S} {U\left( \textbf{s}^c \right){Z^w}\left( \textbf{s}^c \right)},
	\end{equation}
	where $U$ denotes a square window centered at $\textbf{s}^c$, and
	\begin{equation}\nonumber
		U\left( \textbf{s}^c \right) = \exp \left\{ { - \frac{{{{\left[ {{Z^w}\left( \textbf{s}^c \right) - Z_0^w} \right]}^2}}}{{2{\sigma_z ^2}}}} \right\},
	\end{equation}
	where $Z^w_0$ denotes the depth of the closest valid point to $\textbf{s}^c$, the standard deviation $\sigma_z$ is set to 3. So a dense and aligned depth map can be obtained as shown in Fig.~\ref{FIG:FigDenseDepthMap}(a). 
	
	Finally, we apply a median filter with a window size of $3\times3$ followed by a side window box filter~\cite{yin2019side} with a window size of $11\times11$ and 10 iterations. This approach effectively mitigates errors without excessively smoothing the depth values in regions of discontinuity. The resulting reference depth map is presented in Fig.~\ref{FIG:FigDenseDepthMap}(b).
	
\section{Accurate 3D reconstruction} 
\subsection{Image Refocusing}
\begin{figure}[!h]
		\centering
		\includegraphics[width=0.7\linewidth]{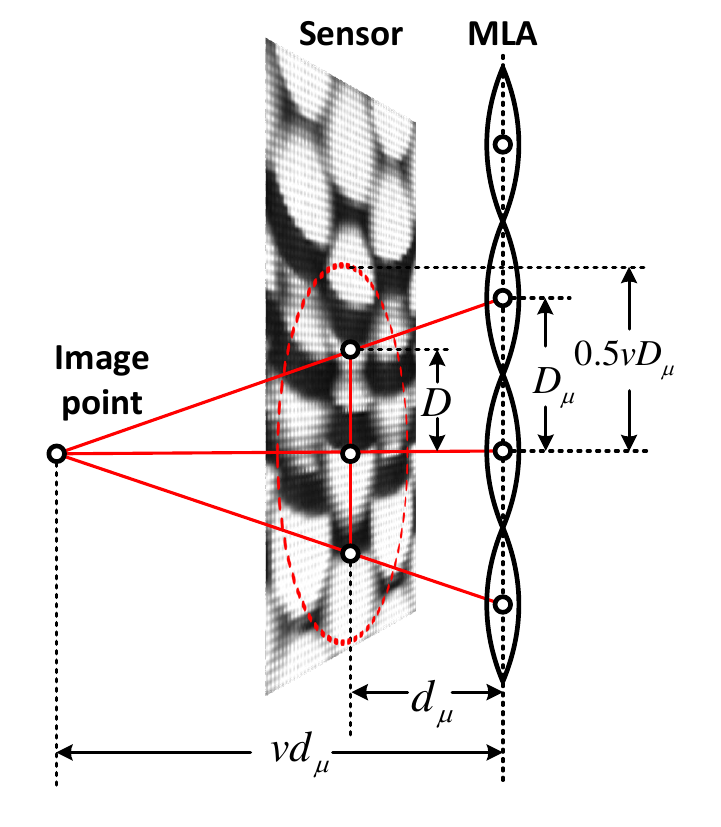}
		\caption{Schematic diagram of refocusing~\cite{heinze2015automated}.} 
		\label{FIG:FigRefocus}
	\end{figure}

	By using the analytical relation between depth and corresponding point distance which can be found in the supplementary material, we convert the reference depth map to a dense and aligned corresponding point distance map, thereby refocusing the captured fringe images. First, we approximately compute the radius of the intersection circle (red dashed circle in Fig.~(\ref{FIG:FigRefocus}) of the MLA plane and the cone of imaging rays as
	\begin{equation}
		r_0^c = 0.5vD_\mu = \frac{0.5D_\mu^2}{D_{\mu}-D}.
	\end{equation}
	Therefore, we can identify the lenslets which are involved with the cone of imaging rays, as they satisfy
	\begin{equation}
		\label{EQ:EqInvolvedLenslet}
		\sqrt{\left(x^c - \alpha^c\right)^2+\left(y^c - \beta^c\right)^2}<r_0^c,
	\end{equation}
	where $(\alpha^c,\beta^c)$ is the pixel coordinate of lenslet image center. For a arbitary pixel $\emph{\textbf{x}}$, its correspondence points are located along the epipolar line with a distance of 
	\begin{equation}
		D'=D'_\mu\left(1-\frac{1}{v}\right),
	\end{equation}
	where $D'_\mu$ is the pixel distance between lenslet centers of $\emph{\textbf{x}}$ and its correspondence point. Finally, we compute the mean value of the intensities for all the corresponding points to have the grayscale value in the refocused image.

 \subsection{Phase Unwrapping and Point Clouds Computing}
\begin{figure}[!h]
		\centering
		\subfloat[]
		{
			\centering
			\includegraphics[width=0.45\linewidth]{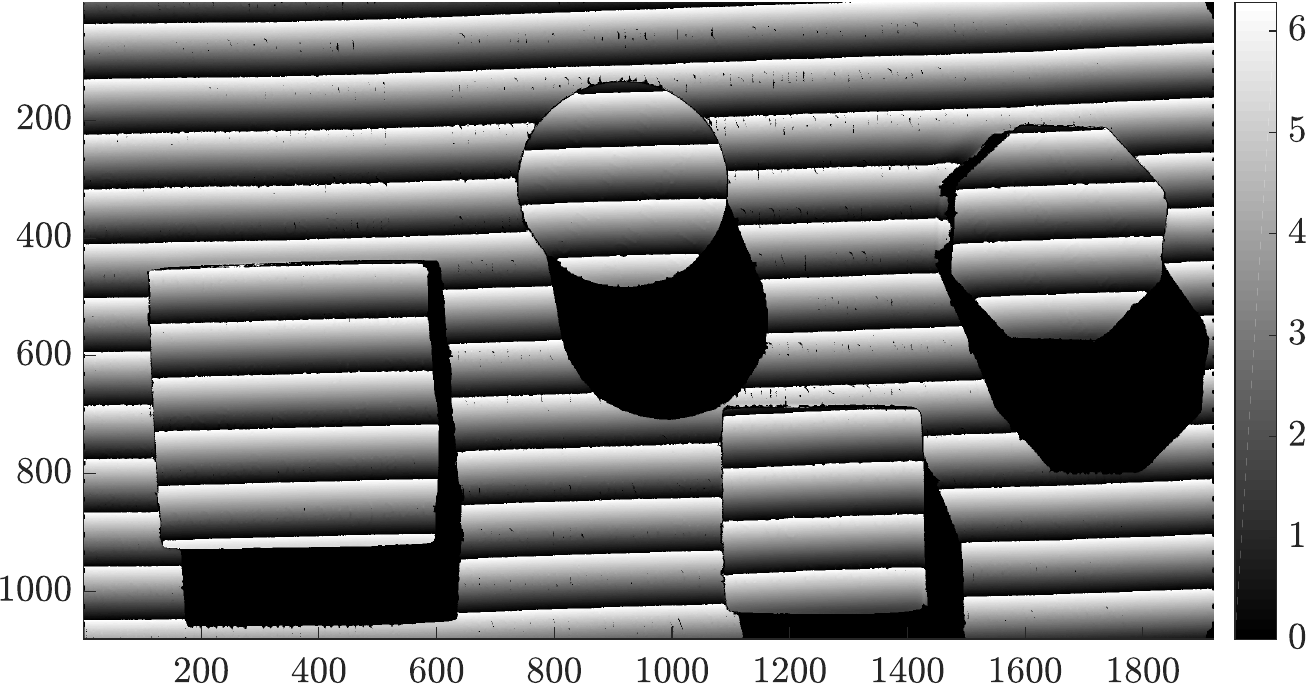}
		}
		\subfloat[]
		{
			\centering
			\includegraphics[width=0.45\linewidth]{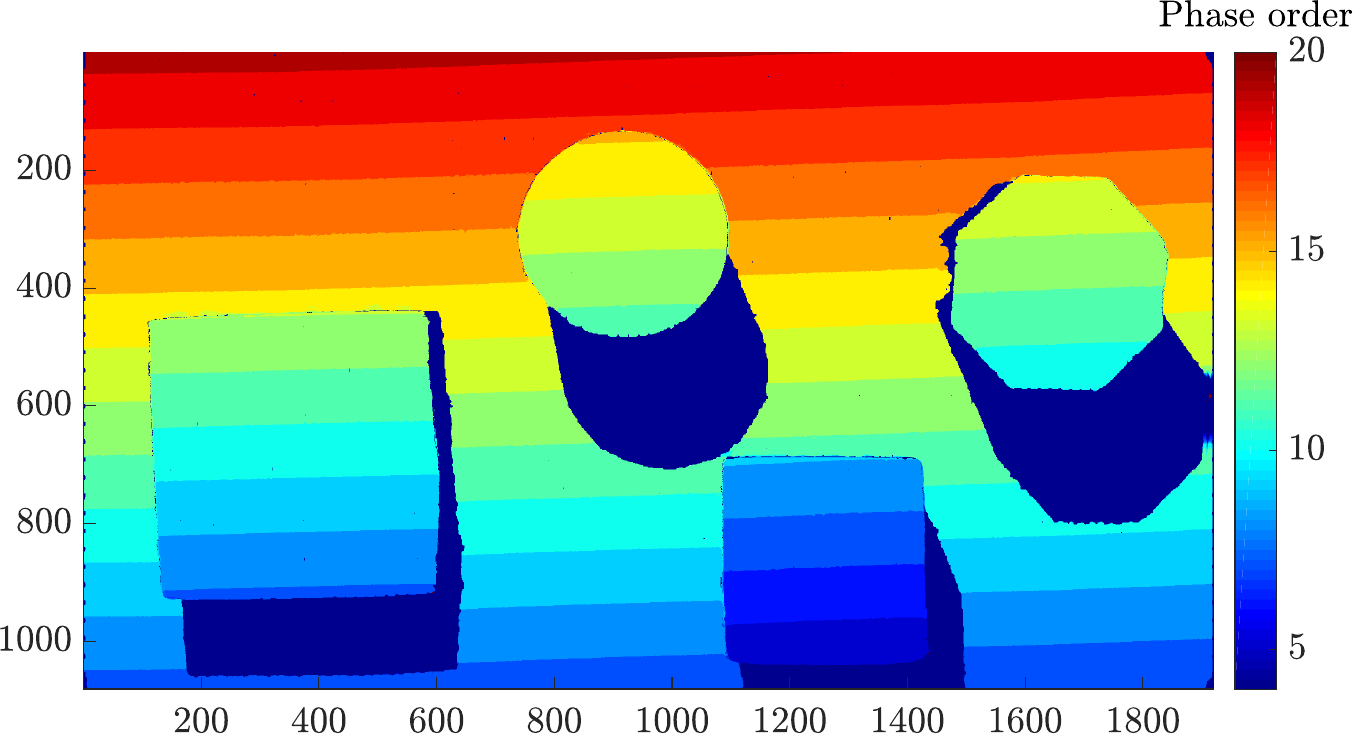}
		}
		\caption{(a) Wrapped phase and (b) phase order computed from Eq.~(\ref{EQ:PhaseUnwrap}).} 
		\label{FIG:FigPhaseUnwrap}
	\end{figure}

    By using the analytical relation between depth and corresponding point distance, we convert the reference depth map to a dense and aligned corresponding point distance map, thereby refocusing the captured fringe images. Implementation details can be found in the supplementary material. From the refocused images, we retrieve the high frequency phase to be unwrapped as shown in Fig.~\ref{FIG:FigPhaseUnwrap}(a). The phase order can be determined by the reference depth map as 
	\begin{equation}\label{EQ:PhaseUnwrap}
		{k_0} = \text{round}\left[\frac{{\left( {m_{21}^pX_r^w + m_{22}^pY_r^w + m_{23}^pZ_r^w + m_{24}^p} \right)f}}{{\left( {m_{31}^pX_r^w + m_{32}^pY_r^w + m_{33}^pZ_r^w + m_{34}^p} \right){H^p}}}\right],
	\end{equation}
	and is visualized in Fig.~\ref{FIG:FigPhaseUnwrap}(b). Thus we compute the absolute phase as
	\begin{equation}
		\Phi  = \frac{\phi  + 2\pi k_0}{f},
	\end{equation}
	
\begin{figure}[!h]
		\centering
		\subfloat[]
		{
			\centering
			\includegraphics[width=0.5\linewidth]{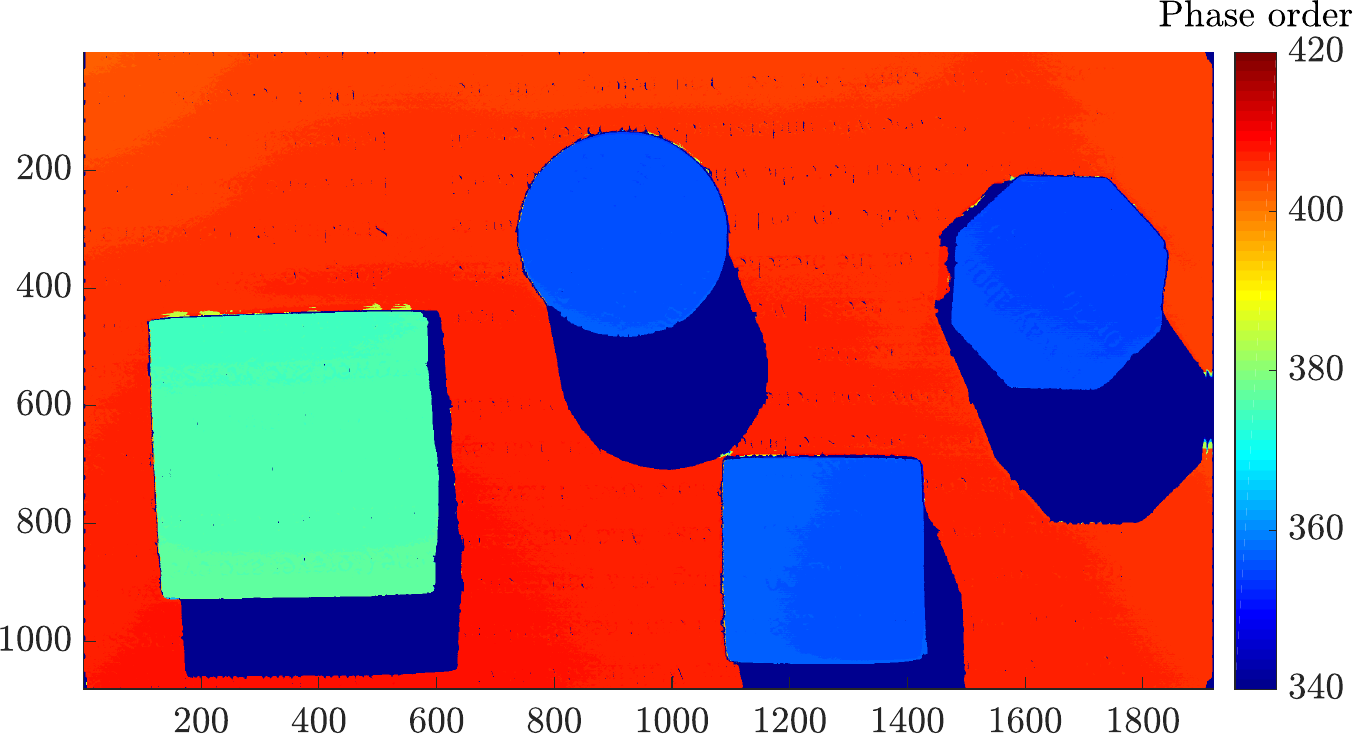}
		}
		\subfloat[]
		{
			\centering
			\includegraphics[width=0.38\linewidth]{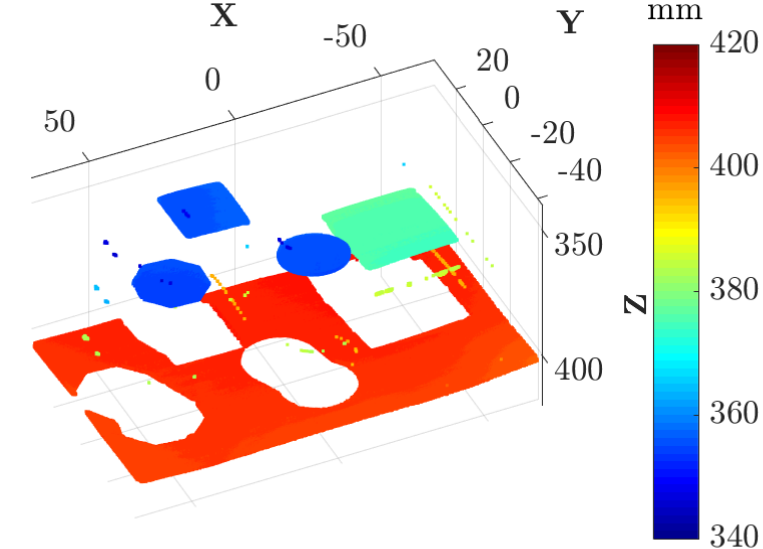}
		}
		\caption{Visualization of reconstructed (a) depth map and (b) point clouds.} 
		\label{FIG:FigAccuratePointCloud}
	\end{figure}

	Finally, the point clouds can be reconstructed by typical triangulation-based algorithm~\cite{liu2010dual,liu2019reconstructing,liu2020extending} as shown in Fig.~\ref{FIG:FigAccuratePointCloud}. Compared with existing ALF 3D imaging techniques, our method significantly improved the resolution of the depth map by an order of magnitude. However, although the depth discontinuous errors in the reference point clouds are effectively reduced by utilizing our PSAD in section 3.2, there still inevitably exists a small number of outliers around the depth jumping regions.
\end{appendices}
\bibliographystyle{splncs04}
\bibliography{main}